\documentclass[acmsmall,screen]{acmart}

\usepackage{graphicx}
\usepackage{textcomp}
\usepackage{xspace}
\usepackage{multirow}
\usepackage{rotating}
\usepackage{tikz}
\usepackage{soul}
\usepackage{tabularx}
\usepackage{url}
\usepackage{pifont}
\usepackage{paralist}
\usepackage{ifthen}
\usepackage[dvipsnames,table]{xcolor}
\usepackage[export]{adjustbox}
\usepackage{amsmath,amsfonts}
\usepackage{listings}
\usepackage{algorithmic}
\usepackage{xspace}
\usepackage{xstring}
\usepackage{ifxetex}
\usepackage{pgfplots}
\pgfplotsset{compat=1.18}
\usepackage{pgfplotstable}
\usetikzlibrary{patterns}
\usepackage{stfloats}
\usepackage[most]{tcolorbox}
\usepackage{dashrule} 
\usepackage{colortbl,xcolor}
\definecolor{headergray}{RGB}{240,240,240}
\definecolor{javablue}{RGB}{210,228,255}   
\definecolor{pyteal}{RGB}{208,243,234}     

\usepackage{xfp}
\usepackage{siunitx}

\usepackage{pgfplots}
\pgfplotsset{compat=1.18}
\definecolor{BaseC}{HTML}{4C78A8}
\definecolor{StructC}{HTML}{B279A2}
\definecolor{SelfCC}{HTML}{59A14F}
\definecolor{ReasonC}{HTML}{F28E2B}
\definecolor{EdgeC}{HTML}{E15759}
\newcommand{\mname}[1]{\textbf{\sffamily #1}}

\newcommand{\pname}[1]{\textbf{\sffamily #1}} 
\newcommand{\bestJ}[1]{\cellcolor{javablue}\textbf{#1}}
\newcommand{\bestP}[1]{\cellcolor{pyteal}\textbf{#1}}

\ifxetex
  \usepackage{fontspec}
  \usepackage{xeCJK}
\fi

\def\BibTeX{{\rm B\kern-.05em{\sc i\kern-.025em b}\kern-.08em
    T\kern-.1667em\lower.7ex\hbox{E}\kern-.125emX}}

\definecolor{darkgreen}{rgb}{0.0, 0.5, 0.0}

\newcommand{\ie}{\emph{i.e.,}\xspace}
\newcommand{\eg}{\emph{e.g.,}\xspace}

\newcommand{\etal}{\emph{et~al.}\xspace}
\newcommand{\secref}[1]{Section~\ref{#1}\xspace}

\newcommand{\figref}[1]{Fig.~\ref{#1}\xspace}

\newcommand{\tabref}[1]{Table~\ref{#1}\xspace}

\newcommand{\clms}{\textit{CLMs}\xspace}
\newcommand{\clmsIn}{Instruction-based \textit{CLMs}\xspace}

\newcommand{\lmsIn}{Instruction-based \textit{LLMs}\xspace}

\newcommand{\shotPrompt}[1]{%
  \IfStrEq{#1}{retrieval}{\textit{3-shot retrieval-based}\xspace}{%
    \IfStrEq{#1}{fixed}{\textit{3-shot fixed}\xspace}{%
      \textit{3-shot #1}\xspace%
    }%
  }%
}

\newcommand{\pPrompt}[2]{$P^{\textsc{#1}}_{\text{\scriptsize #2}}$}

\newcommand{\pBase}[1]{\pPrompt{Base}{#1}}
\newcommand{\pStruct}[1]{\pPrompt{Struct}{#1}}
\newcommand{\pRobust}[1]{\pPrompt{Robust}{#1}}
\newcommand{\pReason}[1]{\pPrompt{Reason}{#1}}

\newcommand{\pEdge}[1]{\pPrompt{Edge}{#1}}

\newcommand{\pXX}[1]{\pPrompt{XX}{#1}}

\newcommand{\uparrowbetter}{\textcolor{ForestGreen}{$\blacktriangle$}}     
\newcommand{\downarrowworse}{\textcolor{BrickRed}{$\blacktriangledown$}}   

\newcommand{\better}[1]{#1~\uparrowbetter}
\newcommand{\worse}[1]{#1~\downarrowworse}

\tcbset{
  on line,
  boxsep=1pt,
  left=2pt,
  right=2pt,
  top=1pt,
  bottom=1pt,
  colframe=blue!70!black, 
  colback=blue!70!black,  
  coltext=white,          
  rounded corners,
  boxrule=0pt             
}

\renewcommand{\rq}[1]{RQ\textsubscript{#1}}

\newcommand{\OR}[1]{\num{\fpeval{1/(#1)}}} 
\newcommand{\pv}[1]{{#1}} 
\newcommand{\na}{\rule{1cm}{5pt}} 

\newcommand{\passat}[2]{\texttt{\text{Pass@}#1\,|\,T{=}#2}}

\newboolean{showcomments}
\setboolean{showcomments}{true}

\ifthenelse{\boolean{showcomments}}{
  \newcommand{\nb}[2]{\fbox{\bfseries\sffamily\scriptsize#1}
    {\sf\small$\blacktriangleright$\textit{#2}$\blacktriangleleft$}
  }
}{
  \newcommand{\nb}[2]{}
}

\newcommand{\stepbadge}[1]{%
\tikz[baseline=(n.base)]\node (n)[
  circle,
  fill=blue!80,
  text=white,
  font=\sffamily\bfseries\small,
  inner sep=1.2pt,
  minimum size=1.3em
]{#1};%
}

\newtcolorbox{promptbox}{colback=white, arc=0.5mm, top=1mm, bottom=1mm, left=1mm, right=1mm, title=System prompt used for generation}
\newtcolorbox{resultbox}{colback=white, arc=0.5mm, top=1mm, bottom=1mm, left=1mm, right=1mm}

\AtBeginDocument{\providecommand\BibTeX{{Bib\TeX}}}

\begin{document}

\title{An Empirical Study on the Effects of System Prompts in Instruction-Tuned Models for Code Generation}

\author{Zaiyu Cheng}
\email{zcheng06@wm.edu}
\affiliation{%
	\institution{AURA @ Dept. of Computer Science, William \& Mary}
	\country{USA}
}

\author{Antonio Mastropaolo}
\email{amastropaolo@wm.edu}
\affiliation{%
	\institution{AURA @ Dept. of Computer Science, William \& Mary}
	\country{USA}
}

\renewcommand{\shortauthors}{Zaiyu et al.}

\begin{abstract}
Instruction-tuned Language Models (\textit{ILMs}) have become essential components of modern AI systems, demonstrating exceptional versatility across natural language and reasoning tasks. Among their most impactful applications is code generation, where \textit{ILMs} -- commonly referred to as Code Language Models (\textit{CLMs}) -- translate human intent into executable programs. While progress has been driven by advances in scaling and training methodologies, one critical aspect remains underexplored: the impact of system prompts on both general-purpose \textit{ILMs} and specialized \textit{CLMs} for code generation. We systematically evaluate how system prompts of varying instructional detail, along with model scale, prompting strategy, and programming language, affect code assistant. Our experimental setting spans \textbf{360} configurations across four models, five system prompts, three prompting strategies, two languages, and two temperature settings. We find that: \textbf{(1)} increasing system-prompt constraint specificity does not monotonically improve correctness—prompt effectiveness is configuration-dependent and can help or hinder based on alignment with task requirements and decoding context; \textbf{(2)} for larger code-specialized models, few-shot examples can degrade performance relative to zero-shot generation, contrary to conventional wisdom; and \textbf{(3)} programming language matters, with Java exhibiting significantly greater sensitivity to system prompt variations than Python, suggesting language-specific prompt engineering strategies may be necessary.
\end{abstract}




\maketitle

\section{Introduction}
\label{sec:intro}
In recent years, instruction tuning has emerged as a critical technique for enhancing the ability of general-purpose Large Language Models (LLMs) to follow specific instructions and interact effectively across diverse domains~\cite{ouyang2022training,wei2021finetuned,shengyu2023instruction,zeng2024automatic}. Building on this foundation, researchers have increasingly extended instruction tuning to Code Language Models (\clms), using instruction-based datasets tailored for software engineering tasks to improve their ability to interpret and execute programming-specific instructions~\cite{wang2023codet5+,luo2023wizardcoder,wei2023magicoder,li-etal-2024-instructcoder,wang-etal-2021-codet5}.

Existing studies in this area primarily focus on optimizing the user prompt component through various techniques and example quantities, which have demonstrated significant performance improvements across code generation, summarization, and translation tasks~\cite{shin2025prompt,feng2024genetic,della2025prompt,Ye2025prompt,khojah2025impact,Brown2020language}. Widely adopted approaches include \textit{zero-shot} prompting, which relies solely on the model's pre-trained knowledge, and \textit{few-shot} prompting, which augments input with demonstrations to guide task adaptation~\cite{kojima2022large,wei2021finetuned}. More advanced strategies include Chain-of-Thought (CoT) reasoning and Retrieval-Augmented Generation (RAG), both proven effective for code generation~\cite{wei2022chain,lewis2020retrieval}.

CoT prompting elicits intermediate reasoning steps that decompose complex problems into manageable sub-tasks, enabling multi-step logical inference before code generation~\cite{kojima2022large,wei2022chain,wang2022self}. In contrast, RAG dynamically retrieves relevant code snippets or natural-language examples from external corpora based on semantic similarity, grounding generation in concrete, contextually relevant examples—particularly beneficial for domain-specific APIs, idiomatic patterns, or project-specific conventions~\cite{lewis2020retrieval,borgeaud2022improving,gao2023retrieval}. However, the role of system-level instructions—which establish the model's behavioral context and operational constraints—remains largely unexplored in the code generation domain.

Beyond prompt strategy, the underlying model architecture and capacity also influence how effectively instructions and examples are interpreted. \clms differ substantially in their pretraining corpora, tokenization schemes, and the balance between general and code-specific objectives. Larger models generally exhibit stronger generalization and reasoning capabilities, yet their behavior may also be more sensitive to system-level instructions and prompt formatting, leading to performance fluctuations when prompt semantics change~\cite{zhuo2024prosa}. In contrast, smaller or mid-sized models, though limited in capacity, often show more predictable responses under consistent prompting setups~\cite{sclar2023quantifying,geng2025control,chatterjee2024posix,sun-etal-2024-enhancing-code}.

Understanding how model scale and specialization interact with system prompts and prompting strategies is essential for three reasons. First, it addresses a methodological gap: existing benchmarks seldom control for system prompt variations, confounding model comparisons and hindering reproducibility \cite{alzahrani2024benchmarks,biderman2024lessons}. Second, it provides practical guidance for practitioners balancing model size, prompting strategy, and instruction design under resource constraints~\cite{sinha2024small}. Third, it offers theoretical insight into whether code-specialized models internalize programming conventions differently from general-purpose models, affecting their responsiveness to explicit instructions versus implicit examples.

\textit{To bridge this gap, we conduct a comparative empirical study of two model families with contrasting specialization strategies: (i)~\textit{code-specialized models} from the Qwen2.5-Coder family~\cite{hui2024qwen2}, trained specifically on code and technical natural language (\ie code documentation), and (ii)~a \textit{general-purpose instruction-tuned model}, GPT-OSS-20B~\cite{openai2025gptoss120bgptoss20bmodel}, trained on broader corpora including -- but not limited to code. While both families share instruction-following capabilities, they diverge fundamentally in their training objectives and domain exposure--\textbf{enabling us to isolate whether prompt sensitivity is shaped by specialized pre-training or is a more universal property of instruction-tuned architectures.}}

The experimental setup addresses a fundamental question: \textit{How robust are LLM-generated solutions to variations in task communication?} Understanding prompt sensitivity is critical because: (1)~developers need reliable code generation \textit{today} under diverse communication styles, not future models~\cite{mastropaolo2023robustness}, (2)~prompt brittleness directly impacts developer productivity and trust in AI-assisted workflows, and (3)~the relative impact of prompt engineering may persist or amplify as models become more instruction-sensitive.

Our experimental design systematically manipulates three dimensions to isolate the impact of prompt formulation on generation quality:

\begin{itemize}
    \item  We evaluate code in \textit{Java} and \textit{Python}--the dominant languages in enterprise and open-source ecosystems--ensuring comparability with established benchmarks~\cite{yu2024codereval, qing2025effibench, cao2024javabench}.

    \item We examine three prompting approaches: one \textit{zero-shot configuration} that supplies only the task description, and two \textit{$n$-shot configurations} that enrich the specification with demonstration examples provided to the model in either static or adaptive form. Each approach is assessed across four models spanning both domain-specialized code architectures and general-purpose instruction-following systems.

    \item We evaluate three deployment scenarios reflecting practical constraints. \textit{Deterministic generation} ($T = 0$, Pass@1) uses greedy decoding to produce the single highest-probability solution, testing top-ranked correctness. \textit{Stochastic single-shot} ($T = 1$, Pass@1) samples from the probability distribution~\cite{li2024eagle,li2025exploring,ouyang2024temperature} but accepts only the first output, measuring reliability under minimal sampling budget. \textit{Stochastic multi-shot} ($T = 1$, Pass@5) generates five candidates, evaluating whether prompt formulation affects solution \textit{diversity} and \textit{coverage} when developers can select among alternatives.

\end{itemize}

To the best of our knowledge, ours is the first study to provide a controlled, surgical analysis of the \textit{interaction} between system prompt design and model characteristics across multiple temperature regimes and sampling budgets, revealing whether prompt sensitivity is a universal phenomenon or varies systematically with model type, scale and/or programming language.
To summarize, we make the following threefold contributions:
\begin{itemize}

\item \textbf{First systematic insights into system-prompt effects in instruction-tuned models for code generation.}
We conduct the first large-scale empirical study examining how system-level instructions influence the behavior of instruction-tuned models in code-generation tasks.

\item \textbf{Uncovering the interplay between general-purpose and code-specialized domains.}
We analyze how model specialization--from general-purpose to code-tuned architectures--modulates responsiveness to system prompts and example-based guidance. 

\item \textbf{A code-generation testbed for system prompt analysis.}
We release a reproducible testbed that standardizes evaluation under controlled variations of system prompts and prompting strategies, including (i) a suite of system instructions, (ii) a retrieval corpus with a fixed three-shot selection policy, (iii) configuration files, seeds, and execution scripts for Java and Python tasks.
\end{itemize}
\section{Background and Related Work}

This section reviews prior studies on \clms and system prompt engineering. We summarize recent progress in \clms for code-related tasks and survey emerging research on the impact of system-level prompts on model behavior and performance.

\subsection{Large Code Models}
Recent advances in adapting LLMs for code-related tasks have primarily focused on instruction tuning and parameter-efficient fine-tuning (PEFT). Yuan \etal~\cite{yuan2023evaluating} systematically evaluated the performance of \textit{ILMs} on code comprehension and generation tasks. The study found that these models demonstrate outstanding capabilities under zero-shot conditions, even outperforming certain task-specific fine-tuned models. Li \etal~\cite{li-etal-2024-instructcoder} proposed \textit{InstructCoder} and constructed a large-scale instruction dataset for code editing to fine-tune models. Experiments showed that the fine-tuned models significantly outperformed baseline models in tasks such as comment insertion, refactoring, and optimization. Luo \etal~\cite{luo2023wizardcoder} introduced \textit{WizardCoder}, which employed the Evol-Instruct method to generate diverse instruction data for model tuning. Results demonstrated substantial improvements on code generation benchmarks such as HumanEval~\cite{chen2021codex} and MBPP~\cite{austin2021program}.

Complementary to instruction tuning, a growing body of research explores PEFT techniques for code models. Chen \etal~\cite{chen2023pass} introduced \textit{Pass-Tuning}, a structure-aware approach that incorporates graph neural representations of abstract syntax trees into prefix tuning to enhance code understanding. Choi and Lee~\cite{choi2023codeprompt} proposed \textit{CodePrompt}, a task-agnostic prefix tuning framework that leverages input-dependent templates and corpus-specific prefixes to bridge the gap between pretraining and downstream adaptation. Fakih \etal~\cite{fakih2024llm4plc} developed \textit{LLM4PLC}, which harnesses LLMs for verifiable programming of Programmable Logic Controllers (PLCs) in industrial control systems, integrating verification tools and user feedback into prompt design. Hajipour \etal~\cite{hajipour2022simscood} presented \textit{SimSCOOD}, a systematic benchmark for analyzing out-of-distribution generalization in fine-tuned source code models across length, syntax, and semantic shifts. Liu \etal~\cite{liu2024mftcoder} proposed \textit{MFTCoder}, a multitask fine-tuning framework that jointly trains models on diverse code-related tasks to improve generalization and training efficiency. Wang \etal~\cite{wang2025teaching} introduced a framework that taught code LLMs to effectively use autocompletion tools during repository-level generation, improving dependency coverage and syntactic validity. Finally, Yang \etal~\cite{yang2024corda} proposed \textit{CorDA}, a context-oriented decomposition adaptation method that employed task-aware low-rank decomposition to enhance parameter-efficient tuning and mitigate catastrophic forgetting of world knowledge.

Beyond adaptation methods, recent progress has also been driven by open foundation code models and the large-scale curation of code-centric pretraining corpora. Rozi\`ere \etal~\cite{roziere2023code} introduced \textit{Code Llama}, a family of open-code models designed for code synthesis and related programming tasks. Lozhkov \etal~\cite{lozhkov2024starcoder} reported \textit{StarCoder2} alongside \textit{The Stack v2}, emphasizing principled data sourcing and extensive coverage of programming languages and software-development artifacts. Guo \etal~\cite{guo2024deepseek} proposed \textit{DeepSeek-Coder}, trained on large-scale code-heavy corpora with objectives targeting project-level completion and infilling.

In parallel, the community has begun to treat instruction data construction itself as a first-class research problem for code models. Wei \etal~\cite{wei2023magicoder} introduced \textit{Magicoder} and the \textit{OSS-Instruct} methodology, which uses open-source code as grounded references to generate more realistic and controllable code instructions for training. Ahmad \etal~\cite{ahmad2025opencodeinstruct} released \textit{OpenCodeInstruct}, a large-scale instruction tuning dataset intended to broaden coverage of code tasks and improve the reliability of supervised instruction tuning for code LLMs. Weyssow \etal~\cite{weyssow2024codeultrafeedback} presented \textit{CodeUltraFeedback}, adopting an LLM-as-a-judge scheme to annotate coding responses with preference-aware feedback, thereby enabling alignment-style training objectives that target user-facing coding preferences beyond pure functional correctness.

Finally, evaluation methodology has evolved to address known gaps in functional-correctness benchmarking. Liu \etal~\cite{liu2023your} proposed \textit{EvalPlus}, which augments canonical programming benchmarks with improved tests to reduce overestimation of correctness and to better stress generalization. Jimenez \etal~\cite{jimenez2023swe} introduced \textit{SWE-bench}, shifting evaluation toward real-world repository settings where models must resolve issues and produce changes that satisfy project tests. These benchmarks underscore that progress on code LLMs increasingly depends not only on model architecture and training, but also on evaluation regimes that more accurately reflect execution-driven correctness and software-engineering realism.

\subsection{System Prompt Engineering}
System prompts are high-level instructions provided to instruction-tuned coder models that define their global behavior across tasks. See \figref{fig:system-prompts} for reference. Unlike user prompts or in-context examples, which are task-specific and can vary with each query, system prompts set overarching guidelines for how the model interprets and generates code. Building on this understanding, recent research has begun to systematically investigate how system-level instructions can be automatically optimized.
Recent analyses further suggest that system prompts form a fragile but critical control surface: as system messages accumulate more guardrails and constraints, models may omit relevant clauses or inconsistently apply higher-priority requirements, motivating dedicated evaluation protocols and interventions tailored to system-prompt adherence~\cite{mu2025closer}.

Zhang \etal~\cite{zhang2024sprig} proposed \textit{SPRIG}, an edit-based genetic algorithm that iteratively constructs system prompts from predefined components. Experiments on $47$ task types showed that a single optimized system prompt can perform on par with task-specific prompts, and combining system-level with task-level optimization can further improve performance. Choi \etal~\cite{choi2025system} introduced \textit{MetaSPO}, framing system prompt optimization as a bilevel problem and solving it via meta-learning. Across $14$ unseen datasets, \textit{MetaSPO} produced system prompts that generalize well under diverse user prompts and support rapid adaptation to new tasks.
Beyond optimizing a single best system prompt, system messages have also been studied as an explicit conditioning channel that can encode preference-like directives at the system level. Lee \etal~\cite{lee2024aligning} proposed training and evaluation protocols that vary the system message itself and examined generalization to diverse, unseen system messages, highlighting that real deployments may require robustness to a distribution of system prompts rather than a single fixed template.

A second thread studies reliability and security implications of system prompts under the instruction hierarchy used by modern assistants. Wallace \etal~\cite{wallace2024instruction} proposed training objectives and data generation methods that teach models to prioritize privileged instructions when higher- and lower-priority directives conflict. Liu \etal~\cite{liu2023prompt} analyzed prompt-injection attacks against LLM-integrated applications, demonstrating that adversarial inputs can manipulate downstream behavior when instruction boundaries are not robustly enforced.
Complementary to integrity attacks that attempt to override system-level intent, recent work has shown that the confidentiality of system prompts is itself a first-class security concern. Hui \etal~\cite{hui2024pleak} proposed \textit{PLeak}, a black-box optimization framework that crafts adversarial queries to induce an LLM application to reveal its hidden system prompt. Levin \etal~\cite{levin2025has} studied prompt membership inference for system prompts, showing that an adversary can statistically test whether a particular system prompt was used, even when direct extraction is difficult. Moving toward structural mitigation, Cao \etal~\cite{cao2025you} proposed encoding system prompts as internal representation vectors rather than raw text in the context, reducing leakage exposure while preserving the behavioral effect of system instructions.
System prompts can also be compromised persistently through attacks that target the system prompt itself or the model supply chain. Li \etal~\cite{li2025system} introduced \textit{system prompt poisoning}, where attackers inject malicious content into the system prompt such that its impact persists across subsequent interactions. Yan \etal~\cite{yan2025system} demonstrated \textit{system prompt hijacking} via permutation triggers embedded in upstream models, enabling downstream bypass of deployer-provided system prompts and exposing a supply-chain vulnerability in distributed LLM development.

\section{Study Design}
The main goal of this study is to systematically investigate how system prompts interact with code generation processes when carried out using both general-purpose instruction-tuned models and code-specialized language models. We aim to decouple and evaluate the individual and combined effects of four key dimensions on model performance: system prompt specificity, model scale, prompting strategy, and programming language. 
To help us study the role of system prompts in shaping code generation outcomes, we pose the following research questions:

\begin{itemize}
\item \textbf{\textit{\rq{1}: How do system prompts with varying levels of specificity affect general-purpose instruction-tuned models?}}
In \rq{1}, we focus on how varying degrees of system prompt specificity affect model behavior across three prompting strategies. Through this analysis, we aim to evaluate the model’s sensitivity to system-level guidance and uncover how prompt granularity impacts the correctness, consistency, and structural organization of the generated code.

\item \textbf{\textit{\rq{2}: How do system prompts with varying levels of specificity affect code-specialized instruction-tuned models?}} In \rq{2}, we extend the analysis conducted in \rq{1}, with the key distinction that the underlying model is explicitly trained to support code-related activities. As in the previous research question, we compare outcomes across different system prompts and prompting strategies to identify potential behavioral differences. Furthermore, we conduct a model-scale analysis within each prompting strategy to examine whether--and to what extent--contextual information and model size interact in shaping the effectiveness of system-level instructions.

\end{itemize}

\begin{figure}[t]
\centering
\resizebox{0.7\linewidth}{!}{%
\begin{tcolorbox}[
    colback=white,        
    colframe=black,      
    title=System prompts,
    coltext=black, 
    fonttitle=\bfseries,
    sharp corners,        
    boxrule=0.8pt,        
    left=4mm,          
    right=4mm,
    top=2mm,
    bottom=2mm,
]
\textbf{Prompt 1 (Baseline):} \\
\quad "You are a highly skilled code generator. Your task is to generate an executable method from the natural language description." \\[1mm]

\textbf{Prompt 2 (Structure-Constrained):} \\
\quad "You are a highly skilled code generator. Your task is to generate an executable method from the natural language description. 

Rules: 

1. Strictly adhere to the function signature, parameter requirements, and output type specified in the docstring or leading comments." \\[1mm]

\textbf{Prompt 3 (Robust-Handling):} \\
\quad "You are a highly skilled code generator. Your task is to generate an executable method from the natural language description. 

Rules: 

1. Strictly adhere to the function signature, parameter requirements, and output type specified in the docstring or leading comments.

2. The method implementation must handle potential invalid inputs and runtime issues with exception-handling behavior as appropriate." \\[1mm]

\textbf{Prompt 4 (Reasoning-Guided):} \\
\quad "You are a highly skilled code generator. Your task is to generate an executable method from the natural language description.

Rules: 

1. Strictly adhere to the function signature, parameter requirements, and output type specified in the docstring or leading comments. 

2. The method implementation must handle potential invalid inputs and runtime issues with exception-handling behavior as appropriate. 

3. Before writing any code, carefully think step by step the method's purpose stated in the docstring or leading comments, and keep this reasoning private." \\[1mm]

\textbf{Prompt 5 (Edge-Coverage):} \\
\quad "You are a highly skilled code generator. Your task is to generate an executable method from the natural language description. 

Rules: 

1. Strictly adhere to the function signature, parameter requirements, and output type specified in the docstring or leading comments. 

2. The method implementation must handle potential invalid inputs and runtime issues with exception-handling behavior as appropriate. 

3. Before writing any code, carefully think step by step the method's purpose stated in the docstring or leading comments, and keep this reasoning private. 

4. The method implementation must handle sufficient edge cases to pass all potential unit tests." \\[1mm]
\end{tcolorbox}
}
\caption{Five system prompts were used in this study.}
\label{fig:system-prompts}

\end{figure}

\subsection{Prompting Strategies}
\label{sec:prompting-strategy}
We structure prompting along two dimensions: (i) \emph{shot availability} and (ii) \emph{prompting technique}. 
For the former, we consider two settings: \emph{zero-shot}, which supplies no examples and relies solely on the model's pretraining, and \emph{few-shot} with a fixed $n=3$ input--output demonstrations per task. 
The rationale for this choice is twofold: ($i$) it aligns with the current state of practice, as evidenced by several recent studies that adopt a similar prompt design strategy ~\cite{xu2024does,pornprasit2024gpt,li2023codeie,yoo2024perc}; and ($ii$) it reflects computational constraints inherent to our experimental setup. Specifically, as detailed in \secref{sec:results}, a total of \textbf{360} model configurations were executed. Consequently, we adopt a number of shots that previous research has shown to provide models with sufficient task-specific patterns for effective adaptation while maintaining prompt conciseness. This design strikes a careful balance between providing adequate contextual information during generation.

The prompting techniques are further divided into two approaches:
\shotPrompt{fixed} prompting that employs a fixed set of general examples uniformly applied across tasks, independent of the input content. These examples remain consistent across queries, offering straightforward guidance without contextual adaptation.

\noindent \shotPrompt{retrieval} prompting dynamically selects examples that most closely resemble the user's input, that is, the requirements the model is asked to implement. This selection is carried out using a third-party dataset or corpus composed of paired instances $\langle \text{Requirement}, \text{Implementing}_{\text{code}} \rangle$, which represent high-quality examples of natural language specifications and their corresponding code implementations. By retrieving and incorporating the most relevant pairs into the prompt, this approach provides contextually aligned demonstrations that can guide the model toward generating more accurate and requirement-consistent code outputs. To retrieve the sought information, we employed a dense embedding retrieval mechanism to identify the most semantically relevant support examples for each query instance~\cite{gao2023precise}. Both the support instance corpus and query code are encoded using the SentenceTransformer model \textit{all-MiniLM-L6-v2}, which has been employed in several research studies when supporting the same retrieval operation \cite{bauerfeind2025david,sheokand2025codemixbench,li2025your,rahman2025relative,mastropaolo2024evaluating,wang2020minilm}. To isolate the effect of prompt design from retrieval-specific variability, we keep the retriever and its hyperparameters fixed throughout all experiments. The normalized embeddings are compared via cosine similarity, and the top three nearest neighbors are selected. The retrieved examples are then integrated into the user prompt.

\subsection{System Prompts Design}
\label{sec:system-prompts}

The evaluation involves multiple system prompt variants that differ in wording, emphasis, and task specificity, while maintaining the underlying style template unchanged. Specifically, as shown in \figref{fig:system-prompts}, system prompts are divided into five variants--\texttt{Prompt 1} (\texttt{Baseline}), \texttt{Prompt 2} (\texttt{Structure-Constrained}), \texttt{Prompt 3} (\texttt{Robust-Handling}), \texttt{Prompt 4} (\texttt{Reasoning-Guided}), and \texttt{Prompt 5} (\texttt{Edge-Coverage)}--forming a progressive sequence in which each prompt introduces distinct levels of constraint or emphasis, allowing us to examine how incremental variations in prompt framing influence model behavior.

The design of these system prompts is informed by human judgment, incorporating thoughtful qualitative reasoning and intentional decisions about task framing and understanding. In particular, these prompts have been designed with structured progression in mind--in both content depth and constraint specificity. As illustrated in \figref{fig:system-prompts}, we vary the \emph{system prompt} along five progressively more constrained variants whose lengths are $21$, $48$, $69$, $100$, and $117$ tokens, respectively, evidencing a monotonic increase in instructional detail and information density. 
Here, \emph{prompt length} refers to the total token count of the system-prompt text for each configuration (excluding the user prompt and any few-shot examples).
Concretely, we define:

\begin{itemize}
    \item \texttt{Baseline}: a minimal configuration with a generic expert role and task description, used to examine the model’s default behavior when generating executable methods solely from natural language specifications.
    \item \texttt{Structure-Constrained}: builds upon the baseline by explicitly enforcing strict adherence to the function signature, parameter list, and return type specified in the docstring, ensuring consistency between the generated implementation and the provided interface.
    \item \texttt{Robust-Handling}: further strengthens the specification by requiring the implementation to handle invalid inputs and runtime issues through appropriate exception-handling mechanisms, encouraging safer and more defensive method design.
    
    \item \texttt{Reasoning-Guided}: extends the robust handling configuration by explicitly activating chain-of-thought internal reasoning. Before emitting any code, the model is instructed to privately work through the stepwise logic implied by the docstring so that the final output remains code only but is informed by an explicit, structured reasoning process.
    
    \item \texttt{Edge-Coverage}: a fully constrained configuration emphasizing correctness and reliability, requiring implementations to account for a broad range of boundary conditions so that the generated methods are more likely to pass potential unit tests, representing the highest level of instruction precision and execution control within this series.
\end{itemize}

\begin{figure}[t]
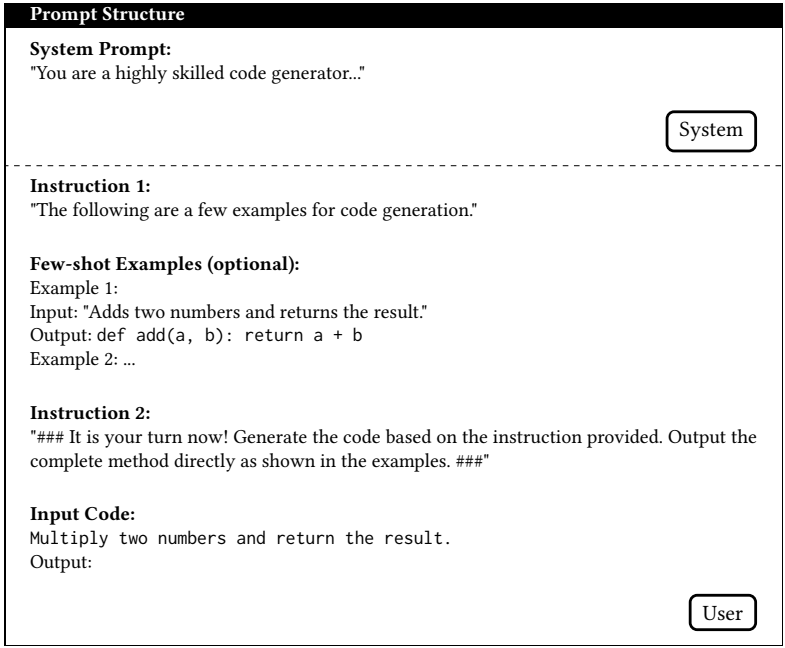

\centering
\resizebox{0.75\linewidth}{!}{%
\begin{tcolorbox}[
    colback=white,
    colframe=black,
    title=Prompt Structure,
    coltext=black,
    fonttitle=\bfseries,
    sharp corners,
    boxrule=0.8pt,
    left=4mm,
    right=4mm,
    top=2mm,
    bottom=2mm
]
\textbf{System Prompt:} \\
\quad "You are a highly skilled code generator..." \\[1mm]

\hfill\tcbox[colframe=black, colback=white, fontupper=\large, nobeforeafter, left=1mm, right=1mm, top=0.5mm, bottom=0.5mm]{System}
\tcblower

\textbf{Instruction 1:} \\
\quad "The following are a few examples for code generation." \\[1mm]

\textbf{Few-shot Examples (optional):} \\
\quad Example 1: \\
\quad \quad Input: "Adds two numbers and returns the result." \\
\quad \quad Output: \texttt{def add(a, b): return a + b} \\
\quad Example 2: ... \\[1mm]

\textbf{Instruction 2:} \\
\quad "\#\#\# It is your turn now! Generate the code based on the instruction provided. Output the complete method directly as shown in the examples. \#\#\#" \\[1mm]

\textbf{Input Code:} \\
\quad \texttt{Multiply two numbers and return the result.} \\
\quad \quad Output: \\

\hfill\tcbox[colframe=black, colback=white, fontupper=\large, nobeforeafter, left=1mm, right=1mm, top=0.5mm, bottom=0.5mm]{User}
\end{tcolorbox}
}
\caption{Representation of the structured prompt used in this study.}
\label{fig:prompt}
\end{figure}

Complementing \figref{fig:system-prompts}, \figref{fig:prompt} depicts the overall two-part prompt template used in our study: (i) the \texttt{System Prompt}, instantiated by one of the five variants above (thereby controlling instruction density and constraints), and (ii) the \texttt{User Prompt}, which contains the task-specific instruction and, when applicable, a few-shot section (fixed at $n{=}3$ demonstrations).

\subsection{Datasets and Evaluation Metrics}
We benchmark each model's configuration against CoderEval \cite{yu2024codereval}--one of the most recent benchmarks for code generation widely employed in related research \cite{crupi2025effectiveness,zhang2025they,wang2025beyond,tambon2025bugs}. It comprises 460 code generation problems evenly distributed across Java (230 problems) and Python (230 problems). Each problem is described by a triplet $\langle d, f, t \rangle$ where: \textbf{(i)} $d$ is a natural language specification describing the requirements for a function to be implemented; \textbf{(ii)} $f$ is a reference implementation demonstrating a correct solution; and \textbf{(iii)} $t$ is a test suite for validating the correctness of automatically generated code.

Utilizing CoderEval as the foundation for measuring system prompt impact in code generation provides a controlled experimental setting in which multiple dimensions--such as those we propose to investigate in Section~\ref{sec:prompting-strategy}--can be systematically examined. Specifically, CoderEval's dual-language structure enables cross-language analysis, allowing us to explore how system prompts interact with code generation tasks across both Python and Java contexts while holding other variables constant, such as task complexity.

The knowledge base (KB) for \shotPrompt{retrieval}  prompting comprises 200,000 examples, each pairing a natural language instruction with its corresponding code implementation. These examples are evenly distributed across the two languages, with 100,000 Java and 100,000 Python samples.

Both retrieval corpora are drawn from the official \textit{CodeXGLUE} benchmark and fall under its text-to-code scenario; we therefore use the language-aligned resources provided by \textit{CodeXGLUE}--CONCODE~\cite{lu2021codexglue} for Java and CodeSearchNet AdvTest~\cite{diera2023gencodesearchnet} for Python—while applying identical retrieval and preprocessing procedures across languages to reduce potential corpus-induced confounding.


Following standard practice in code generation research \cite{chen2021codex,li2024evocodebench}, we employ the \textbf{Pass@k} metric to evaluate model performance, which measures the proportion of generated solutions that pass all unit tests among the top $k$ candidates. In this study, we set $k=1$ and $k=5$, reporting both \textbf{Pass@1} and \textbf{Pass@5} because they capture complementary aspects of model capability. \textbf{Pass@1} measures the success rate under a single deterministic generation, which reflects the experience of users interacting with a coding assistant or automated code generation system in realistic single-attempt scenarios. \textbf{Pass@5} allows multiple independent samples and evaluates whether a correct solution exists among several plausible candidates, which reveals the model's potential performance when search or reranking strategies are available. Together, these two metrics distinguish immediate reliability from latent solution diversity. Formally, if a test set contains $N$ problems, and $C$ of the model's top-1 outputs are correct, Pass@1 is computed as:

\[
\text{Pass@1} = \frac{C}{N} \times 100\%
\]

In addition, if a test set contains \(N\) problems and for each problem \(i\) we generate five candidate solutions, let \(s_i\) be an indicator that equals one if at least one of the five candidates passes all tests and equals zero otherwise. Pass@5 is computed as:

\[
\text{Pass@5} = \frac{\sum_{i=1}^{N} s_i}{N} \times 100\%
\]

To test whether more information-dense system prompts outperform the \texttt{Base} prompt, we apply McNemar’s test for paired binary outcomes~\cite{mcnemar1947note}. Each code-generation instance is treated as a matched observation evaluated under \texttt{Baseline} and, respectively, \texttt{Structure-Constrained}, \texttt{Robust-Handling}, \texttt{Reasoning-Guided}, and \texttt{Edge-Coverage} prompts (success = Pass@k is \textit{True}, failure = Pass@k is \textit{False}). We report the resulting $p$-values alongside matched-pairs Odds Ratios (ORs) and 95\% confidence intervals to quantify effect magnitude and direction. Finally, to account for multiple comparisons, we correct p-values using \textit{Holm correction}~\cite{yates1934contingency}.



\subsection{Instruction-Tuned Model Selection}

To conduct our experiments, we rely on two types of instruction-tuned models that represent complementary approaches to LLM development and specialization. The first category includes general-purpose instruction-tuned models designed to follow natural language instructions across diverse tasks. The second category comprises domain-specialized variants, particularly models exposed to large-scale code datasets to enhance code understanding. By doing so, we further investigate the generalization capacity of broadly instruction-tuned models and the specialized capabilities of code-oriented models under varying prompting conditions.

\begin{itemize}

\item \textbf{GPT-OSS}~\cite{openai2025gptoss120bgptoss20bmodel}: A general-purpose, open-source, instruction-tuned model family comprising 20B and 120B variants. These models are optimized for multi-domain reasoning and demonstrate performance comparable to proprietary instruction-tuned systems such as Claude~\cite{anthropic_claude3_modelcard_2024} and GPT-4~\cite{achiam2023gpt} across a range of code generation benchmarks~\cite{jimenez2024swebench,quan2025codeelo}. Consequently, GPT-OSS offers an effective balance between openness, scalability, and performance--delivering capabilities that closely align with those of widely adopted closed-source models like GPT-4, while maintaining full transparency and reproducibility. This makes it a particularly suitable choice for examining how system-level prompts affect the behavior of general-purpose, instruction-tuned models in code generation tasks. In this study, we focus on GPT-OSS-20B as the representative model for our experiments.

\item \textbf{Qwen2.5-Coder}~\cite{hui2024qwen2}: A domain-specialized model family trained on large-scale code corpora and supporting a diverse range of programming languages and instruction formats. The Qwen2.5-Coder series is well established in the field of software engineering automation and represents the current state of practice for benchmarking code generation models \cite{wang2025co,ashraf2025toward,liu2025rstar,wang2024seed}. In this study, we examine several variants of Qwen2.5-Coder, specifically the 1.5B, 7B, and 32B parameter configurations. This allows us to systematically analyze how system-level prompts interact with models of varying scales within a domain-specialized, code-oriented family.

\end{itemize}

\subsection{Model Inference}
\label{sec:model-inference}
Each experimental task used the official \textbf{Hugging Face Transformers} implementation of the evaluated models\footnote{\url{https://huggingface.co/collections/Qwen/qwen25-coder}}\textsuperscript{,}\footnote{\url{https://huggingface.co/openai/gpt-oss-20b}}. For every configuration--defined by the combination of model type, model scale, programming language, prompting strategy, inference temperature, sampling parameters, and system prompt--we conducted a separate inference run. Each prompt variant was evaluated independently to avoid potential cross-contamination of context or caching effects across runs. In total, this procedure yielded \textbf{360 distinct model configurations}, as outlined in Section~\ref{sec:system-prompts}.

As anticipated in \secref{sec:intro}, we evaluate both \clmsIn and \lmsIn under two different temperature regimes: $T=0$ and $T=1$.
Comparing these two settings enables us to analyze performance differences between deterministic decoding ($T=0$) and realistic deployment scenarios ($T=1$), leading to a more robust and externally valid assessment of model behavior.

During the post-processing phase of the generated outputs, we observed that several responses contained formatting artifacts, such as language delimiters (\eg \texttt{java} or \texttt{python}) preceding the code. These artifacts often caused compilation or execution failures during metric evaluation, which could erroneously reduce Pass@1 scores to zero. To mitigate this, we implemented a preprocessing step that systematically removed these language-specific tokens and extraneous text segments, ensuring that all generated code could be correctly parsed and executed in subsequent testing stages.

\subsection{Experimental Environment }
All experiments are conducted on a workstation running \textbf{Ubuntu 24.04.3 LTS} with the \textbf{GNU/Linux 6.8.0-85-generic} kernel. The system is equipped with four \textbf{NVIDIA L40S} GPUs, each with \textbf{48 GB} of VRAM. All model inferences in this study are performed using \textbf{vLLM}.


\section{Results}
\label{sec:results}
In this section, we present and discuss the results of our investigation, structured according to the research questions outlined earlier.  For readability, we adopt several naming conventions throughout this paper. 

\begin{itemize}
\item \textbf{\textit{Model names}}: We abbreviate models in the Qwen2.5-Coder family and GPT-OSS-20B by retaining only their parameter scale. For example, Qwen2.5-Coder-1.5B becomes Qwen-1.5B, and GPT-OSS-20B becomes GPT-20B.\\

\item \textbf{\textit{System Prompts}}:  We reference prompts using the numeric identifiers from \secref{sec:system-prompts}: \texttt{Prompt 1} corresponds to Base (\figref{fig:system-prompts}), \texttt{Prompt 2} to Structural, continuing through \texttt{Prompt 5}.\\

\item  \textbf{\textit{Language Specific Prompts}}:  When reporting results across programming languages, we use \pXX{Java} and \pXX{Python} to denote prompt-language pairings (\eg \pBase{Java} indicates the Base prompt applied to Java). \\

\item \textbf{\textit{Experimental Configurations}}: We denote experimental setups using the following notation: \passat{k}{T}, where $k$ is the number of generated samples and $T$ is the sampling temperature. For instance, \passat{1}{1} indicates Pass@1 evaluation using code generated at temperature $T=1$.
\end{itemize}

\subsection{\rq{1}:  \textit{How do system prompts with varying levels of specificity affect general-purpose instruction-tuned models?}}
\label{sec:rq1}
This section is organized into three subsections that correspond to our decoding and evaluation regimes: \textbf{RQ1.1} (\passat{1}{0}), \textbf{RQ1.2} (\passat{1}{1}), and \textbf{RQ1.3} (\passat{5}{1}). Within each subsection, we contrast the three prompting strategies used in this study -- \textit{zero-shot}, \textit{3-shot fixed}, and \textit{3-shot retrieval-based} prompting.

\smallskip

\noindent Before presenting the results, we briefly summarize the table structure used throughout \secref{sec:rq1}. Each table reports results for a single backbone model, GPT-OSS. Within every table, we organize results into two language-specific blocks: Java (top) and Python (bottom). Each row corresponds to an evaluated model configuration, while the columns compare the \pname{Base} prompt against progressively more information-dense variants (\pname{Struct}, \pname{Robust}, \pname{Reason}, and \pname{Edge}).
For each prompt variant, we report: (i) Pass@1 accuracy (\%), (ii) the $p$-value from McNemar's test computed relative to \pname{Base}, and (iii) the Odds Ratio (OR) capturing effect size and direction. The symbols \textcolor{ForestGreen}{$\blacktriangle$} and \textcolor{BrickRed}{$\blacktriangledown$} denote improvements and degradations relative to \pname{Base}, respectively; “–” indicates no change; and black bars mark non-significant differences ($p \ge 0.05$).

\subsubsection{\textbf{RQ1.1: \passat{1}{0}}}
As shown in Table~\ref{tab:gpt-pass@1-t0}, GPT-20B performs best when the system prompt emphasizes structural requirements, regardless of the target language. In Java, \pStruct{Java} achieves 42.17\% Pass@1, exceeding \pBase{Java} (39.57\%) and \pRobust{Java} (37.83\%). Python exhibits the same trend: \pStruct{Python} yields 24.35\% functionally correct solutions compared to 20.87\% for \pBase{Python}, whereas \pEdge{Python} drops to 16.96\%, suggesting that edge-coverage prompts can distract from core task constraints when no examples are provided.

However, these apparent differences are not statistically significant: McNemar's tests against \pname{Base} do not reject the null hypothesis at the 0.05 level (as indicated by the black bars), implying that the observed gains and drops may reflect sampling variability rather than a reliable effect of the prompting variant.

Introducing examples through \shotPrompt{fixed} prompting  alters model behavior. For Java, GPT-20B exhibits marked vulnerability to example context under minimal guidance: \pBase{Java} plummets to 14.35\% Pass@1. Performance recovers incrementally as prompt complexity increases, rising to 20.43\% with \pStruct{Java}, 25.22\% with \pRobust{Java}, 26.09\% with \pReason{Java}, and peaking at 29.57\% with \pEdge{Java}. Even this peak remains well below the 42.17\% \textit{zero-shot} baseline, indicating that introducing fixed examples undermines Java performance on the whole, with richer prompts only partially mitigating—rather than reversing—the harm from suboptimal example choice. These differences achieve statistical significance across all pairwise comparisons. Indeed, progressively elaborate system prompts yield ORs ranging from $2.17$ to $5.38$, meaning that under optimal conditions (\pEdge{Java}), the model has $5 \times$ higher odds of generating functionally correct code than under \pBase{Java}.

Python demonstrates a contrasting pattern: \pBase{Python} and \pEdge{Python} tie at 22.17\%, while \pStruct{Python} and \pRobust{Python} are lower at 19.13\% and 19.57\%. However, these differences are not statistically significant, suggesting only a directional tendency that added constraints do not help when demonstrations are present.

\begin{table*}[t]
\centering
\small
\renewcommand{\arraystretch}{1.2}
\setlength{\tabcolsep}{4pt}
\arrayrulecolor{black}
\resizebox{\textwidth}{!}{%
{\color{black}
\begin{tabular}{|l|c|ccc|ccc|ccc|ccc|}
\toprule
\rowcolor{headergray}
\textbf{Model}
& \multicolumn{13}{c|}{\textbf{GPT-OSS-20B}} \\
\rowcolor{headergray}

& \multicolumn{1}{c|}{\pname{Base}}
& \multicolumn{3}{c|}{\pname{Struct}}
& \multicolumn{3}{c|}{\pname{Robust}}
& \multicolumn{3}{c|}{\pname{Reason}}
& \multicolumn{3}{c|}{\pname{Edge}} \\
\rowcolor{headergray}
\textbf{Strategy}
& \multicolumn{1}{c|}{\textbf{Pass@1}}
& \textbf{Pass@1} & \textbf{$p$-value} & \textbf{OR}
& \textbf{Pass@1} & \textbf{$p$-value} & \textbf{OR}
& \textbf{Pass@1} & \textbf{$p$-value} & \textbf{OR}
& \textbf{Pass@1} & \textbf{$p$-value} & \textbf{OR} \\
\midrule

\multicolumn{14}{c}{\textbf{Java}} \\
\midrule

\mname{Zero-shot}
& 39.57
& \better{\bestJ{42.17}} & \na & \na
& \worse{37.83} & \na & \na
& \better{40.43} & \na & \na
& \better{40.00} & \na & \na \\

\mname{Fixed}
& 14.35
& \better{20.43} & \pv{$<0.05$} & \OR{0.4615}
& \better{25.22} & \pv{$<0.05$} & \OR{0.2857}
& \better{26.09} & \pv{$<0.05$} & \OR{0.25}
& \better{\bestJ{29.57}} & \pv{$<0.05$} & \OR{0.186} \\

\mname{Retrieval-based}
& 26.09
& \better{36.96} & \pv{$<0.05$} & \OR{0.1935}
& \better{39.13} & \pv{$<0.05$} & \OR{0.2105}
& \better{\bestJ{40.87}} & \pv{$<0.05$} & \OR{0.0811}
& \better{39.57} & \pv{$<0.05$} & \OR{0.1143} \\
\midrule

\multicolumn{14}{c}{\textbf{Python}} \\
\midrule
\mname{Zero-shot}
& 20.87
& \better{\bestP{24.35}} & \na & \na
& \worse{20.43} & \na & \na
& \better{22.61} & \na & \na
& \worse{16.96} & \na & \na \\

\mname{Fixed}
& \bestP{22.17}
& \worse{19.13} & \na & \na
& \worse{19.57} & \na & \na
& \worse{20.87} & \na & \na
& \bestP{22.17} -- & \na & \na \\

\mname{Retrieval-based}
& 20.87
& 20.87 -- & \na & \na
& \worse{20.43} & \na & \na
& \better{\bestP{22.17}} & \na & \na
& \worse{20.00} & \na & \na \\
\bottomrule
\end{tabular}
}}
\vspace{10pt}
\caption{\textbf{Performance results for GPT-OSS-20B, evaluated with \passat{1}{0}} }
\label{tab:gpt-pass@1-t0}
\end{table*}

\textit{3-shot retrieval-based} prompting partially resolves the instability introduced by fixed examples, especially for Java. Starting from 26.09\% under \pBase{Java}, GPT-20B improves to 36.96\% with \pStruct{Java} and 39.13\% with \pRobust{Java}, reaching its best \textit{retrieval-based} score of 40.87\% under \pReason{Java}. This configuration comes within 1.30\% of the overall best result for Java, the \textit{zero-shot} \pStruct{Java} score of 42.17\%, suggesting that carefully selected examples can recover most of the loss introduced by fixed examples, but only when paired with a prompt that steers the model toward deliberate requirement satisfaction. The performance gains observed when enriching the user prompt with contextually relevant code generation examples yield statistically significant improvements across all system prompts examined in this study, with ORs spanning from $4.75$ to $12.33$.

For Python, \textit{retrieval-based} prompting produces a considerably tighter performance range of 20.00\% to 22.17\%, with \pReason{Python} again achieving the highest score at 22.17\%. Unlike Java, retrieval enhances consistency but does not unlock substantial performance gains, confirming that GPT-20B's Python code generation is less responsive to system prompt variations when examples are relevant, though overall correctness remains limited. This distinctive behavior is further reflected in our statistical analyses, which failed to detect significant differences across conditions.

Across languages, in most cases, GPT-20B achieves higher peak performance on Java than on Python for the same evaluation protocol. Under \textit{zero-shot}, the best Java configuration exceeds the best Python configuration by 17.82\%, and under \textit{retrieval-based} the corresponding gap is 18.70\%. The exception is \shotPrompt{fixed} prompting, where the gap shrinks to 7.40\% because Java degrades dramatically under \pBase{Java} while Python remains comparatively resilient.

\begin{tcolorbox}[
  enhanced,
  breakable,
  colback=cyan!8,        
  colframe=black,        
  coltext=black,
  arc=8pt,               
  boxrule=0.8pt,         
  left=6pt,right=6pt,top=10pt,bottom=8pt,
  fonttitle=\bfseries,
  title=Answer to RQ1.1,   
  coltitle=black,
  varwidth boxed title,  
  attach boxed title to top left={yshift=-3mm,xshift=6mm},
  boxed title style={
    colback=gray!35,     
    colframe=black,      
    boxrule=0.8pt,
    arc=10pt,            
    outer arc=10pt,
    left=6pt,right=6pt,top=2pt,bottom=2pt,
  }
  ]
System prompt complexity influences GPT-20B differently across prompting strategies and languages. Under \textit{zero-shot}, structural prompts (\pStruct{}) achieve highest scores (Java: 42.17\%, Python: 24.35\%) without statistical significance. \shotPrompt{fixed} prompting severely degrades Java (14.35\% to 29.57\%, OR: 2.17--5.38) while Python remains stable (19.13\%--22.17\%). \textit{Retrieval-based} prompting recovers Java performance substantially (40.87\%, OR: 4.75--12.33) but minimally affects Python (20.00\%--22.17\%, non-significant). Overall, Java benefits more from prompt engineering and example quality than Python.
\end{tcolorbox}


\begin{table*}[t]
\centering
\small
\renewcommand{\arraystretch}{1.2}
\setlength{\tabcolsep}{4pt}
\arrayrulecolor{black}
\resizebox{\textwidth}{!}{%
{\color{black}
\begin{tabular}{|l|c|ccc|ccc|ccc|ccc|}
\toprule
\rowcolor{headergray}
\textbf{Model}
& \multicolumn{13}{c|}{\textbf{GPT-OSS-20B}} \\
\rowcolor{headergray}

& \multicolumn{1}{c|}{\pname{Base}}
& \multicolumn{3}{c|}{\pname{Struct}}
& \multicolumn{3}{c|}{\pname{Robust}}
& \multicolumn{3}{c|}{\pname{Reason}}
& \multicolumn{3}{c|}{\pname{Edge}} \\
\rowcolor{headergray}
\textbf{Strategy}
& \multicolumn{1}{c|}{\textbf{Pass@1}}
& \textbf{Pass@1} & \textbf{$p$-value} & \textbf{OR}
& \textbf{Pass@1} & \textbf{$p$-value} & \textbf{OR}
& \textbf{Pass@1} & \textbf{$p$-value} & \textbf{OR}
& \textbf{Pass@1} & \textbf{$p$-value} & \textbf{OR} \\
\midrule

\multicolumn{14}{c}{\textbf{Java}} \\
\midrule

\mname{Zero-shot}
& 37.83
& \better{41.30} & \na & \na
& \better{40.00} & \na & \na
& \better{39.57} & \na & \na
& \better{\bestJ{42.61}} & \na & \na \\

\mname{Fixed}
& 18.26
& \better{25.65} & \pv{$<0.05$} & \OR{0.5278}
& \better{30.87} & \pv{$<0.05$} & \OR{0.2368}
& \better{\bestJ{34.35}} & \pv{$<0.05$} & \OR{0.2128}
& \better{32.17} & \pv{$<0.05$} & \OR{0.2381} \\

\mname{Retrieval-based}
& 27.83
& \better{35.22} & \pv{$<0.05$} & \OR{0.4333}
& \better{36.52} & \pv{$<0.05$} & \OR{0.3548}
& \better{41.30} & \pv{$<0.05$} & \OR{0.2051}
& \better{\bestJ{44.78}} & \pv{$<0.05$} & \OR{0.1333} \\
\midrule

\multicolumn{14}{c}{\textbf{Python}} \\
\midrule
\mname{Zero-shot}
& \bestP{25.65}
& \worse{23.48} & \na & \na
& \worse{24.78} & \na & \na
& \worse{23.48} & \na & \na
& \worse{23.91} & \na & \na \\

\mname{Fixed}
& 24.35
& \better{25.65} & \na & \na
& \better{\bestP{26.09}} & \na & \na
& \better{25.22} & \na & \na
& \better{25.22} & \na & \na \\

\mname{Retrieval-based}
& 21.30
& \better{\bestP{23.04}} & \na & \na
& \better{21.74} & \na & \na
& \worse{20.87} & \na & \na
& \better{\bestP{23.04}} & \na & \na \\
\bottomrule
\end{tabular}
}}
\vspace{10pt}
\caption{\textbf{Performance results for GPT-OSS-20B, evaluated with \passat{1}{1}}}
\label{tab:gpt-pass@1-t1}
\end{table*}
\subsubsection{\textbf{RQ1.2: \passat{1}{1}}}
Table~\ref{tab:gpt-pass@1-t1} reveals that under \textit{zero-shot} prompting, the model demonstrates a clear affinity for more elaborate system-level constraints in Java method generation. Performance climbs from 37.83\% with \pBase{Java} to 41.30\% with \pStruct{Java}, maintains strength at 40.00\% under \pRobust{Java}, and reaches its zenith at 42.61\% with \pEdge{Java}. While these gains are quantitatively observable, McNemar's test indicates they do not achieve statistical significance.

The Python \textit{zero-shot} results show an almost inverted pattern. Here, GPT-20B performs best with the minimalist prompt: \pBase{Python} reaches 25.65\%, while every more articulated variant yields lower accuracy, from 24.78\% under \pRobust{Python} down to 23.48\% under both \pStruct{Python} and \pReason{Python}. This indicates that, for Python, additional instruction content can function as overhead rather than guidance, reasonably because the language imposes fewer language-mandated syntactic requirements, making it easier to enter with default priors, while extra constraints increase the chance of over-steering or misprioritizing what the tests actually enforce.

In contrast to the $T = 0$ setting above, increasing the temperature to one changes how GPT-20B trades off default priors versus system-level guidance, especially when examples are present. Under \shotPrompt{fixed} prompting, adding examples amplifies prompt sensitivity in Java code generation tasks. The same GPT-20B model that achieved 37.83\% with \pBase{Java} under \textit{zero-shot} collapses to 18.26\% when fixed examples are introduced. However, as the system prompt becomes more prescriptive, performance bounces back sharply: \pStruct{Java} reaches 25.65\%, \pRobust{Java} reaches 30.87\%, and \pReason{Java} peaks at 34.35\%, with \pEdge{Java} close behind at 32.17\%. These improvements yield not only quantitative gains but also notable consistency, as evidenced by ORs of 1.89 for \pStruct{Java}, 4.22 for \pRobust{Java}, 4.70 for \pReason{Java}, and 4.20 for \pEdge{Java}. A plausible interpretation is that noisy contextual information necessitates explicit global guidance to help the model distinguish meaningful patterns from artifacts in examples. Reasoning-based instructions steer the model in realigning generation toward functional correctness.

In Python, the \shotPrompt{fixed} setting yields comparatively modest and stable changes. GPT-20B moves from 24.35\% with \pBase{Python} to a best of 26.09\% with \pRobust{Python}, while \pStruct{Python}, \pReason{Python}, and \pEdge{Python} cluster around 25.22\% to 25.65\%. In line with the above conclusion, this smaller fluctuation range indicates that Python generation is less dependent on system-prompt instructions.

Table~\ref{tab:gpt-pass@1-t1} also shows the strongest overall Java result for GPT-20B under \textit{3-shot retrieval-based} prompting, where example relevance is controlled. With progressively stronger constraints, Java accuracy increases approximately linearly, from 27.83\% under \pBase{Java} to 44.78\% under \pEdge{Java}. 
Critically, this best \textit{retrieval-based} configuration surpasses the best \textit{zero-shot} Java approach by 2.17\%, showing that examples become effective when query-aligned and paired with expert-style system prompts. Retrieval thus not only stabilizes but elevates performance, contingent on clear validity specifications in the system prompt. As with fixed-shot selection, statistically significant improvements confirm the value of contextual guidance.

Python contrasts with Java again: \textit{Retrieval-based} prompting does not improve GPT-20B, with the best score only 23.04\% under \pStruct{Python} and \pEdge{Python}, below both the \textit{zero-shot} best of 25.65\% and the \shotPrompt{fixed} prompting best of 26.09\%. Moreover, \pReason{Python} drops to 20.87\%, suggesting that, in Python, coupling retrieved examples with explicit reasoning guidance can lead the model to overfit to contextual patterns or to introduce unnecessary complexity that harms correctness. This reinforces a consistent theme across the tables: Python performance is comparatively robust to prompt variation -- a result that is also supported by the statistical comparison, which has produced for each single configuration $p$-values larger than $0.05$, thus rejecting our original hypothesis.

\begin{tcolorbox}[
  enhanced,
  breakable,
  colback=cyan!8,        
  colframe=black,        
  coltext=black,
  arc=8pt,               
  boxrule=0.8pt,         
  left=6pt,right=6pt,top=10pt,bottom=8pt,
  fonttitle=\bfseries,
  title=Answer to RQ1.2,   
  coltitle=black,
  varwidth boxed title,  
  attach boxed title to top left={yshift=-3mm,xshift=6mm},
  boxed title style={
    colback=gray!35,     
    colframe=black,      
    boxrule=0.8pt,
    arc=10pt,            
    outer arc=10pt,
    left=6pt,right=6pt,top=2pt,bottom=2pt,
  }
  ]
Overall, the variability itself is a key signal. In Java, the Pass@1 range across system prompts is 4.78\% under \textit{zero-shot}, then expands to 16.09\% under \shotPrompt{fixed} and 16.95\% under \textit{3-shot retrieval-based}, revealing that examples substantially increase the importance of system prompt design. In Python, ranges remain tight, never exceeding 2.17\%, implying a weaker dependence on how instructions are phrased. Finally, the cross-language gap changes markedly by strategy: Java exceeds Python by 16.96\% at the best \textit{zero-shot} configurations, the gap shrinks to 8.26\% under \shotPrompt{fixed}, then widens sharply to 21.74\% under \textit{3-shot retrieval-based}. This pattern suggests that GPT-20B is able to exploit relevant contextual evidence far more effectively in Java than in Python, and that example relevance paired with high-information-density instruction disproportionately benefits the Java setting.
\end{tcolorbox}

\subsubsection{\textbf{RQ1.3: \passat{5}{1}}}

When the deployment scenario allows practitioners to inspect multiple code recommendations, we naturally expect improved overall functional correctness. This is precisely what emerges from Table~\ref{tab:gpt-pass@5-t1}.

Under \textit{zero-shot} prompting, both languages exhibit minimal sensitivity to system prompt variations. Java performance clusters tightly between 48.70\% (\pBase{Java}, \pEdge{Java}) and 50.43\% (\pRobust{Java}), while Python ranges from 33.91\% to 34.78\% (\pRobust{Python}). These narrow spreads indicate that GPT-20B can generate correct solutions reliably without aggressive prompt specialization when no examples are provided.

\shotPrompt{fixed} prompting reveals divergent language-specific responses. Java benefits from increased guidance: performance climbs from 45.65\% (\pBase{Java}) to 50.43\% (\pRobust{Java}), with \pEdge{Java} maintaining strength at 47.57\%. This suggests that detailed constraints help filter example noise and maintain alignment with test requirements. Conversely, Python performs best with minimal instruction (\pBase{Python}: 36.09\%), declining to 33.04\% under both \pStruct{Python} and \pEdge{Python}. Here, fixed examples appear to provide sufficient grounding, with additional constraints functioning as friction rather than guidance.


\begin{table*}[t]
\centering
\small
\renewcommand{\arraystretch}{1.2}
\setlength{\tabcolsep}{4pt}
\arrayrulecolor{black}
\resizebox{\textwidth}{!}{%
{\color{black}
\begin{tabular}{|l|c|ccc|ccc|ccc|ccc|}
\toprule
\rowcolor{headergray}
\textbf{Model}
& \multicolumn{13}{c|}{\textbf{GPT-OSS-20B}} \\
\rowcolor{headergray}

& \multicolumn{1}{c|}{\pname{Base}}
& \multicolumn{3}{c|}{\pname{Struct}}
& \multicolumn{3}{c|}{\pname{Robust}}
& \multicolumn{3}{c|}{\pname{Reason}}
& \multicolumn{3}{c|}{\pname{Edge}} \\
\rowcolor{headergray}
\textbf{Strategy}
& \multicolumn{1}{c|}{\textbf{Pass@5}}
& \textbf{Pass@5} & \textbf{$p$-value} & \textbf{OR}
& \textbf{Pass@5} & \textbf{$p$-value} & \textbf{OR}
& \textbf{Pass@5} & \textbf{$p$-value} & \textbf{OR}
& \textbf{Pass@5} & \textbf{$p$-value} & \textbf{OR} \\
\midrule

\multicolumn{14}{c}{\textbf{Java}} \\
\midrule

\mname{Zero-shot}
& 48.70
& \better{49.57} & \na & \na
& \better{\bestJ{50.43}} & \na & \na
& \better{49.57} & \na & \na
& 48.70 -- & \na & \na \\

\mname{Fixed}
& 45.65
& \better{46.52} & \na & \na
& \better{\bestJ{50.43}} & \na & \na
& \better{47.39} & \na & \na
& \better{47.57} & \na & \na \\

\mname{Retrieval-based}
& 46.52
& \better{47.83} & \na & \na
& \better{50.00} & \na & \na
& \better{50.87} & \na & \na
& \better{\bestJ{53.91}} & \pv{$<0.05$} & \OR{0.1053} \\
\midrule

\multicolumn{14}{c}{\textbf{Python}} \\
\midrule
\mname{Zero-shot}
& 33.91
& \better{34.35} & \na & \na
& \better{\bestP{34.78}} & \na & \na
& \better{34.35} & \na & \na
& 33.91 -- & \na & \na \\

\mname{Fixed}
& \bestP{36.09}
& \worse{33.04} & \na & \na
& \worse{34.78} & \na & \na
& \worse{35.65} & \na & \na
& \worse{33.04} & \na & \na \\

\mname{Retrieval-based}
& 33.91
& \better{35.65} & \na & \na
& \better{34.78} & \na & \na
& \better{\bestP{36.52}} & \na & \na
& \better{34.35} & \na & \na \\
\bottomrule
\end{tabular}
}}
\vspace{10pt}
\caption{\textbf{Performance results for GPT-OSS-20B, evaluated with \passat{5}{1}.} }
\label{tab:gpt-pass@5-t1}
\end{table*}

\textit{Retrieval-based} prompting demonstrates the strongest synergy between example quality and prompt sophistication. For Java, \pEdge{Java} achieves the study's highest score of 53.91\%, surpassing both \textit{zero-shot} and \textit{fixed} configurations. This indicates that relevant examples combined with comprehensive constraints maximize GPT-20B's potential. \pReason{Java} (50.87\%) and \pRobust{Java} (50.00\%) remain highly competitive, while simpler prompts like \pBase{Java} (46.52\%) underperform, confirming that retrieval alone is insufficient without adequate system-level guidance.

Python shows modest but meaningful improvements under retrieval, peaking at 36.52\% with \pReason{Python}. Performance now follows a clearer ordering across prompts (35.65\% for \pStruct{Python}, 34.78\% for \pRobust{Python}), suggesting that relevant examples help when paired with reasoning-focused guidance. However, absolute gains remain limited, indicating that Python's correctness barriers stem more from intrinsic generation errors than from insufficient contextual cues. 

Moreover, statistically significant benefits from the more advanced system prompts emerge only in one setting. 
When contextual cues are selected via a code-similarity mechanism, the effect strengthens, reaching an OR of 9.50 -- but only when using the most comprehensive system prompt, \ie \pEdge{Java}, which explicitly specifies the environment, context, and constraints. Python's inherent stability is further validated even under the more dynamic conditions examined in this subsection.

\vspace{10pt}
\begin{tcolorbox}[
  enhanced,
  breakable,
  colback=cyan!8,        
  colframe=black,        
  coltext=black,
  arc=8pt,               
  boxrule=0.8pt,         
  left=6pt,right=6pt,top=10pt,bottom=8pt,
  fonttitle=\bfseries,
  title=Answer to RQ1.3,   
  coltitle=black,
  varwidth boxed title,  
  attach boxed title to top left={yshift=-3mm,xshift=6mm},
  boxed title style={
    colback=gray!35,     
    colframe=black,      
    boxrule=0.8pt,
    arc=10pt,            
    outer arc=10pt,
    left=6pt,right=6pt,top=2pt,bottom=2pt,
  }
]

With \passat{5}{1} sampling, Java consistently outperforms Python in functional correctness. Java is highly sensitive to prompt design once examples are provided, performing best when relevant exemplars are paired with information-rich, constraint-explicit instructions. Python remains comparatively stable across prompt variants and tends to benefit most from simpler guidance that avoids unnecessary constraints, suggesting that Java optimization relies on tight example–prompt alignment, or calibration procedure, while Python does not, and favors minimal corrections.
\end{tcolorbox}

\subsection{RQ2: \textit{How do system prompts with varying levels of specificity affect code-specialized instruction-tuned models?}}

To answer and present the results of RQ2, we follow the same three-subsection structure as in \secref{sec:rq1}.

The difference is in the interpretation of the table. In this section, each table summarizes one prompting strategy, and reports results for a single backbone model—either GPT-OSS or Qwen-Coder. Results are split into two language sections, with Java presented first (top) and Python second (bottom). Rows enumerate the model backbones under evaluation (\eg Qwen-Coder), whereas columns contrast the \pname{Base} prompt with increasingly information-rich variants (\pname{Struct}, \pname{Robust}, \pname{Reason}, and \pname{Edge}).
For each variant, we report: (i) Pass@1, (ii) the
$p$-value from McNemar's test comparing against \pname{Base}, and (iii) the Odds Ratio (OR) as a measure of effect size and direction. Improvements and regressions relative to \pname{Base} are indicated by \textcolor{ForestGreen}{$\blacktriangle$} and \textcolor{BrickRed}{$\blacktriangledown$}, respectively; ``–'' denotes no difference; and black bars highlight results that are not statistically significant ($p \ge 0.05$).

\subsubsection{\textbf{RQ2.1: \passat{1}{0}}} As shown in \tabref{tab:zero-pass@1-t0}, with \textit{zero-shot} prompting, Qwen's Java performance is strongly capacity-dependent: Qwen-32B sets the highest ceiling and benefits from more restrictive guidance (peaking at 42.61\% with \pReason{Java} and 41.74\% with \pRobust{Java}), Qwen-7B improves moderately and consistently over \pBase{Java}, and Qwen-1.5B benefits only from structural guidance (\pStruct{Java}, 35.65\%) while degrading sharply once constraints intensify (\pRobust{Java}–\pEdge{Java}, 26–27\%). Importantly, statistical significance is the exception rather than the rule in this regime: across most model–prompt pairs the McNemar tests do not reject the null (reported as ``\na'' in the table), and the few significant outcomes are concentrated in rare edge cases -- notably, the smallest model (Qwen-1.5B) shows statistically significant degradations when moving from \pBase{Java} to more constraint-heavy prompts, while only one significant improvement appears for Qwen-32B (the \pReason{Java} variant).

Compared to the GPT-20B (last row)-- \tabref{tab:zero-pass@1-t0} -- the pattern is consistent: GPT-20B is already strong in \pBase{Java}: 39.57\%, best at \pStruct{Java}: 42.17\%, but, like Qwen-32B, does not exhibit widespread significant gains across prompt variants in the zero-shot, $T=0$ setting. 

In comparison to Java, the Python results are more compact and admit a clearer notion of an attainable \emph{upper bound} that shifts with model scale. At the top end, Qwen-32B defines the strongest ceiling, with \pStruct{Python} reaching 26.52\% (up from 23.91\% with \pBase{Python}); the remaining variants are tightly clustered, with \pRobust{Python} and \pEdge{Python} both at 24.35\% and \pReason{Python} matching the baseline at 23.91\%. For Qwen-7B, the best configuration shifts toward the most explicit prompt, as \pEdge{Python} attains 22.61\% relative to a 19.57\% baseline, while the other variants remain in a narrow 20.00\%--20.87\% band. Finally, Qwen-1.5B shows the same ``fragility'' as in Java: it gains only slightly with \pStruct{Python} (19.13\% vs.\ 18.70\%) and drops sharply with stricter prompts, falling to 15.22\%--14.78\% from \pRobust{Python} to \pEdge{Python}.

Overall, Python outcomes tighten more rapidly than Java as model capacity increases: prompt specialization mainly nudges the best achievable score for larger models, rather than reshaping the full performance profile. Consistent with this pattern, the statistical tests in \tabref{tab:zero-pass@1-t0} indicate that prompt effects are the exception in \textit{zero-shot} \passat{1}{0}; in Python, differences across prompt variants are uniformly non-significant, underscoring a limited sensitivity to system prompt choice at $T=0$.

Comparing GPT-20B to the largest instruction-tuned code model, \ie Qwen-32B, shows that OpenAI's open-source model reaches its best Python accuracy of 24.35\% with \pStruct{Python}. This score falls below the Qwen-32B upper bound (26.52\%) but exceeds the best Qwen-7B configuration (22.61\%), suggesting that a general-purpose instruction-tuned model can remain competitive despite being optimized for a broader set of tasks beyond, for example, code generation.

More broadly, the GPT-20B row reinforces the main takeaway of \tabref{tab:zero-pass@1-t0}; although point estimates vary across prompt variants (\eg 20.87\%--24.35\% for GPT-20B), statistically significant improvements are rare in the zero-shot $T{=}0$ regime, and most differences are not distinguishable from \pBase{} in either language at $p < 0.05$.

\definecolor{gptossrow}{HTML}{F2F2F2} 


\begin{table*}[t]
\centering
\small
\renewcommand{\arraystretch}{1.2}
\setlength{\tabcolsep}{4pt}
\arrayrulecolor{black}
\resizebox{\textwidth}{!}{%
{\color{black}
\begin{tabular}{|l|c|ccc|ccc|ccc|ccc|}
\toprule
\rowcolor{headergray}
\textbf{Strategy}
& \multicolumn{13}{c|}{\textbf{Zero-shot Prompting}} \\
\rowcolor{headergray}

& \multicolumn{1}{c|}{\pname{Base}}
& \multicolumn{3}{c|}{\pname{Struct}}
& \multicolumn{3}{c|}{\pname{Robust}}
& \multicolumn{3}{c|}{\pname{Reason}}
& \multicolumn{3}{c|}{\pname{Edge}} \\
\rowcolor{headergray}
\textbf{Model}
& \multicolumn{1}{c|}{\textbf{Pass@1}}
& \textbf{Pass@1} & \textbf{$p$-value} & \textbf{OR}
& \textbf{Pass@1} & \textbf{$p$-value} & \textbf{OR}
& \textbf{Pass@1} & \textbf{$p$-value} & \textbf{OR}
& \textbf{Pass@1} & \textbf{$p$-value} & \textbf{OR} \\
\midrule

\multicolumn{14}{c}{\textbf{Java}} \\
\midrule

\mname{Qwen-1.5B}
& 32.61
& \better{\bestJ{35.65}} & \na & \na
& \worse{27.39} & \pv{$<0.05$} & \OR{2.7143}
& \worse{26.96} & \pv{$<0.05$} & \OR{2.8571}
& \worse{26.52} & \pv{$<0.05$} & \OR{3.3333}\\

\mname{Qwen-7B}
& 32.61
& \better{\bestJ{36.52}} & \na & \na
& \better{35.65} & \na & \na
& \better{36.09} & \na & \na
& \better{35.22} & \na & \na \\

\mname{Qwen-32B}
& 37.39
& 37.39 -- & \na & \na
& \better{41.74} & \na & \na
& \better{\bestJ{42.61}} & \pv{$<0.05$} & \OR{0.25}
& \better{39.57} & \na & \na \\
\midrule

\rowcolor{gptossrow}
\mname{\textbf{GPT-20B}}
& 39.57
& \better{\bestJ{42.17}} & \na & \na
& \worse{37.83} & \na & \na
& \better{40.43} & \na & \na
& \better{40.00} & \na & \na \\
\bottomrule
\multicolumn{14}{c}{\textbf{Python}} \\
\midrule

\mname{Qwen-1.5B}
& 18.70
& \better{\bestP{19.13}} & \na & \na
& \worse{15.22} & \na & \na
& \worse{15.65} & \na & \na
& \worse{14.78} & \na & \na \\

\mname{Qwen-7B}
& 19.57
& \better{20.87} & \na & \na
& \better{20.00} & \na & \na
& \better{20.43} & \na & \na
& \better{\bestP{22.61}} & \na & \na \\

\mname{Qwen-32B}
& 23.91
& \better{\bestP{26.52}} & \na & \na
& \better{24.35} & \na & \na
& 23.91 -- & \na & \na
& \better{24.35} & \na & \na \\

\bottomrule

\rowcolor{gptossrow}
\mname{\textbf{GPT-20B}}
& 20.87
& \better{\bestP{24.35}} & \na & \na
& \worse{20.43} & \na & \na
& \better{22.61} & \na & \na
& \worse{16.96} & \na & \na \\
\bottomrule
\end{tabular}
}}
\vspace{10pt}
\caption{\textbf{Performance results for the zero-shot prompting strategy, evaluated with \passat{1}{0}.}}
\label{tab:zero-pass@1-t0}
\end{table*}
Compared with the \textit{zero-shot} setting, \shotPrompt{fixed} prompting markedly reshapes both absolute performance and prompt sensitivity, especially for Java (\tabref{tab:random-pass@1-t0}). For the larger models, fixed examples appear highly disruptive unless counterbalanced by more prescriptive system guidance. In Qwen-32B, \pBase{Java} collapses to 6.96\%, but performance rebounds sharply with richer prompts, reaching 29.57\% with \pRobust{Java} and 27.83\% with \pEdge{Java}. Qwen-7B follows the same trajectory: \pBase{Java} drops to 13.04\%, then improves monotonically as constraints increase, peaking at 28.70\% with \pEdge{Java}. By contrast, Qwen-1.5B is comparatively stable and even benefits from mild structure: it starts at 29.57\% with \pBase{Java} and attains its best score of 35.22\% with \pStruct{Java}, while heavier prompt variants yield smaller gains.

The McNemar results in \tabref{tab:random-pass@1-t0} clarify which of these shifts are reliable. In Java, many of the improvements for Qwen-7B and Qwen-32B (and also GPT-20B) are statistically significant relative to \pBase{Java} (frequently $p < 0.05$), indicating that more contextually driven system prompts consistently mitigate the noise introduced by fixed examples. However, we can anticipate that in the generated Python code, it does not produce any significant effect as we vary the type of system prompt and underlying model size. Overall, fixed examples amplify prompt sensitivity primarily in Java---and the statistical tests confirm that this interaction is both large in magnitude and, in several configurations, statistically reliable.

\definecolor{gptossrow}{HTML}{F2F2F2} 


\begin{table*}[t]
\centering
\small
\renewcommand{\arraystretch}{1.2}
\setlength{\tabcolsep}{4pt}
\arrayrulecolor{black}
\resizebox{\textwidth}{!}{%
{\color{black}
\begin{tabular}{|l|c|ccc|ccc|ccc|ccc|}
\toprule
\rowcolor{headergray}
\textbf{Strategy}
& \multicolumn{13}{c|}{\textbf{Fixed Prompting}} \\
\rowcolor{headergray}

& \multicolumn{1}{c|}{\pname{Base}}
& \multicolumn{3}{c|}{\pname{Struct}}
& \multicolumn{3}{c|}{\pname{Robust}}
& \multicolumn{3}{c|}{\pname{Reason}}
& \multicolumn{3}{c|}{\pname{Edge}} \\
\rowcolor{headergray}
\textbf{Model}
& \multicolumn{1}{c|}{\textbf{Pass@1}}
& \textbf{Pass@1} & \textbf{$p$-value} & \textbf{OR}
& \textbf{Pass@1} & \textbf{$p$-value} & \textbf{OR}
& \textbf{Pass@1} & \textbf{$p$-value} & \textbf{OR}
& \textbf{Pass@1} & \textbf{$p$-value} & \textbf{OR} \\
\midrule

\multicolumn{14}{c}{\textbf{Java}} \\
\midrule


\mname{Qwen-1.5B}
& 29.57
& \better{35.22} & \pv{$<0.05$} & \OR{0.2778}
& \better{32.17} & \na & \na
& \better{32.61} & \na & \na
& \better{\bestJ{33.04}} & \na & \na \\

\mname{Qwen-7B}
& 13.04
& \better{24.78} & \pv{$<0.05$} & \OR{0.10}
& \better{25.65} & \pv{$<0.05$} & \OR{0.0938}
& \better{27.83} & \pv{$<0.05$} & \OR{0.0811}
& \better{\bestJ{28.70}} & \pv{$<0.05$} & \OR{0.1429} \\

\mname{Qwen-32B}
& 6.96
& \better{16.96} & \pv{$<0.05$} & \OR{0.08}
& \better{\bestJ{29.57}} & \pv{$<0.05$} & \OR{0.037}
& \better{22.61} & \pv{$<0.05$} & \OR{0.10}
& \better{27.83} & \pv{$<0.05$} & \OR{0.0769} \\

\midrule
\rowcolor{gptossrow}
\mname{\textbf{GPT-20B}}
& 14.35
& \better{20.43} & \pv{$<0.05$} & \OR{0.4615}
& \better{25.22} & \pv{$<0.05$} & \OR{0.2857}
& \better{26.09} & \pv{$<0.05$} & \OR{0.25}
& \better{\bestJ{29.57}} & \pv{$<0.05$} & \OR{0.186} \\

\midrule
\multicolumn{14}{c}{\textbf{Python}} \\
\midrule


\mname{Qwen-1.5B}
& 17.83
& \better{18.70} & \na & \na
& \better{\bestP{20.00}} & \na & \na
& \better{19.13} & \na & \na
& \better{\bestP{20.00}} & \na & \na \\

\mname{Qwen-7B}
& 20.00
& \worse{19.57} & \na & \na
& 20.00 -- & \na & \na
& \better{20.43} & \na & \na
& \better{\bestP{21.30}} & \na & \na \\

\mname{Qwen-32B}
& 23.04
& \better{\bestP{23.91}} & \na & \na
& \worse{22.17} & \na & \na
& \worse{22.17} & \na & \na
& \better{23.48} & \na & \na \\

\midrule
\rowcolor{gptossrow}
\mname{\textbf{GPT-20B}}
& \bestP{22.17}
& \worse{19.13} & \na & \na
& \worse{19.57} & \na & \na
& \worse{20.87} & \na & \na
& \bestP{22.17} -- & \na & \na \\

\bottomrule
\end{tabular}
}}
\vspace{10pt}
\caption{\textbf{Performance results for the 3-shot fixed prompting strategy, evaluated with \passat{1}{0}.}}
\label{tab:random-pass@1-t0}
\end{table*}
\definecolor{gptossrow}{HTML}{F2F2F2} 


\begin{table*}[t]
\centering
\small
\renewcommand{\arraystretch}{1.2}
\setlength{\tabcolsep}{4pt}
\arrayrulecolor{black}
\resizebox{\textwidth}{!}{%
{\color{black}
\begin{tabular}{|l|c|ccc|ccc|ccc|ccc|}
\toprule
\rowcolor{headergray}
\textbf{Strategy}
& \multicolumn{13}{c|}{\textbf{Retrieval-based Prompting}} \\
\rowcolor{headergray}

& \multicolumn{1}{c|}{\pname{Base}}
& \multicolumn{3}{c|}{\pname{Struct}}
& \multicolumn{3}{c|}{\pname{Robust}}
& \multicolumn{3}{c|}{\pname{Reason}}
& \multicolumn{3}{c|}{\pname{Edge}} \\
\rowcolor{headergray}
\textbf{Model}
& \multicolumn{1}{c|}{\textbf{Pass@1}}
& \textbf{Pass@1} & \textbf{$p$-value} & \textbf{OR}
& \textbf{Pass@1} & \textbf{$p$-value} & \textbf{OR}
& \textbf{Pass@1} & \textbf{$p$-value} & \textbf{OR}
& \textbf{Pass@1} & \textbf{$p$-value} & \textbf{OR} \\
\midrule

\multicolumn{14}{c}{\textbf{Java}} \\
\midrule

\mname{Qwen-1.5B}
& 26.09
& \better{\bestJ{31.74}} & \pv{$<0.05$} & \OR{0.1875}
& \better{28.70} & \na & \na
& \better{30.87} & \na & \na
& \better{31.30} & \na & \na \\

\mname{Qwen-7B}
& 12.17
& \better{15.65} & \pv{$<0.05$} & \OR{0.2727}
& \better{19.57} & \pv{$<0.05$} & \OR{0.2273}
& \better{17.83} & \pv{$<0.05$} & \OR{0.381}
& \better{\bestJ{21.30}} & \pv{$<0.05$} & \OR{0.25} \\

\mname{Qwen-32B}
& 17.83
& \better{30.43} & \pv{$<0.05$} & \OR{0.0938}
& \better{\bestJ{36.09}} & \pv{$<0.05$} & \OR{0.1064}
& \better{34.78} & \pv{$<0.05$} & \OR{0.093}
& \better{34.35} & \pv{$<0.05$} & \OR{0.1556} \\

\midrule
\rowcolor{gptossrow}
\mname{\textbf{GPT-20B}}
& 26.09
& \better{36.96} & \pv{$<0.05$} & \OR{0.1935}
& \better{39.13} & \pv{$<0.05$} & \OR{0.2105}
& \better{\bestJ{40.87}} & \pv{$<0.05$} & \OR{0.0811}
& \better{39.57} & \pv{$<0.05$} & \OR{0.1143} \\

\midrule
\multicolumn{14}{c}{\textbf{Python}} \\
\midrule

\mname{Qwen-1.5B}
& 17.39
& \worse{16.52} & \na & \na
& \worse{16.96} & \na & \na
& \worse{16.96} & \na & \na
& \better{\bestP{17.83}} & \na & \na \\

\mname{Qwen-7B}
& 20.00
& \worse{19.57} & \na & \na
& \better{22.17} & \na & \na
& \better{22.61} & \na & \na
& \better{\bestP{24.35}} & \na & \na \\

\mname{Qwen-32B}
& 25.65
& \better{\bestP{26.09}} & \na & \na
& \worse{25.22} & \na & \na
& \worse{23.91} & \na & \na
& \worse{24.78} & \na & \na \\

\midrule
\rowcolor{gptossrow}
\mname{\textbf{GPT-20B}}
& 20.87
& 20.87 -- & \na & \na
& \worse{20.43} & \na & \na
& \better{\bestP{22.17}} & \na & \na
& \worse{20.00} & \na & \na \\

\bottomrule
\end{tabular}
}}
\vspace{10pt}
\caption{\textbf{Performance results for the 3-shot retrieval-based prompting strategy, evaluated with \passat{1}{0}}}
\label{tab:retrieval-pass@1-t0}
\end{table*}

In Python, retrieval-based prompting stabilizes performance and yields modest gains, most clearly for Qwen-7B, which reaches 24.35\% with \pEdge{Python} (above both its \textit{zero-shot} and fixed best). Qwen-32B remains near its \textit{zero-shot} ceiling, peaking at 26.09\% with \pStruct{Python}, while Qwen-1.5B stays tightly bounded (16.52\%--17.83\%), indicating limited benefit from retrieved context.

Across strategies at $\passat{1}{0}$, Java is markedly more sensitive than Python: fixed examples cause the largest disruptions (notably for Qwen-7B and Qwen-32B), whereas retrieval yields the strongest recovery (especially for Qwen-32B). Prompt preferences also vary with scale, with Qwen-1.5B most consistently helped by \pStruct{}, while larger models shift toward more constraint-heavy prompts in Java. In Python, preferences are comparatively stable, with Qwen-32B favoring \pStruct{Python} and Qwen-7B favoring \pEdge{Python}.

The statistical results in \tabref{tab:retrieval-pass@1-t0} reinforce these performance trends. In Java, McNemar's tests against \pBase{Java} are frequently significant ($p<0.05$) for Qwen-7B, Qwen-32B, and GPT-20B, indicating that retrieval combined with richer system prompts yields consistent gains rather than sampling noise. The corresponding effect sizes are substantial rather than incremental: for GPT-20B, the odds of producing functionally correct code increase by roughly 4.75--12.33$\times$ over \pBase{Java} across the more contextually rich prompt variants, and for a code-specialized backbone such as Qwen-32B, the odds increases remain comparably large (about 6.43--10.75$\times$).

\begin{tcolorbox}[
  enhanced,
  breakable,
  colback=cyan!8,        
  colframe=black,        
  coltext=black,
  arc=8pt,               
  boxrule=0.8pt,         
  left=6pt,right=6pt,top=10pt,bottom=8pt,
  fonttitle=\bfseries,
  title=Answer to RQ2.1,   
  coltitle=black,
  varwidth boxed title,  
  attach boxed title to top left={yshift=-3mm,xshift=6mm},
    boxed title style={
        colback=gray!40,
        boxrule=0.7pt,
        arc=8pt,
        outer arc=8pt,
        left=5pt,
        right=5pt,
        top=0.5pt,
        bottom=0.5pt,
    }
]

This first set of results -- obtained under tightly controlled conditions with $T=0$ to minimize stochasticity and ensure replicability -- yields a coherent narrative. In zero-shot settings, system prompt specificity acts as a secondary factor: model scale and language drive outcomes, with statistically significant differences remaining uncommon. Introducing examples, however, transforms this dynamic. Fixed examples can severely destabilize Java generation for larger models, while retrieval-based examples restore stability and enable substantial gains when paired with more informative system prompts. The shift reveals that example selection method fundamentally alters how models respond to prompt design.

\end{tcolorbox}

\subsubsection{\textbf{RQ2.2: \passat{1}{1}}}
Relative to \textbf{RQ2.1}, increasing the temperature to one systematically hinders the ability of the Qwen family to generate functionally correct code. This temperature shift is most visible in \textit{zero-shot} Java code generation: for example, Qwen-1.5B drops from 32.61\% under \pBase{Java} $T = 0$ to 28.26\% (\tabref{tab:zero-pass@1-t1}), and its earlier best case under \pStruct{Java} at 35.65\% no longer persists. In other words, once token generation shifts from greedy decoding to stochastic sampling, prompt design is no longer only about task specification; it also determines whether the model can maintain functional alignment under sampling noise.

In particular, with \textit{zero-shot} prompting, the larger Qwen backbones (Qwen-7B and especially Qwen-32B) generally gain from richer, more descriptive system prompts, whereas the smallest model (Qwen-1.5B) is consistently affected (negatively) as prompt specificity increases. Notably, for Qwen-1.5B the drop in performance under \pReason{} and \pEdge{} is large enough to yield statistically significant differences relative to \pBase{}, while for the larger models similar prompt-induced shifts rarely reach significance despite comparable directional trends.

General-purpose instruction-tuned models behave somewhat differently, as shown by the last row of \tabref{tab:zero-pass@1-t1}. In particular, GPT-20B achieves its best performance -- which can even surpass a code-specialized model in this setting -- only when guided by a more descriptive system prompt that imposes clearer constraints and thereby narrows the effective search space.

Python reveals a different pattern under zero-shot conditions. Both specialized and general models generate fewer correct solutions, showing selective sensitivity to prompt formulation rather than consistent gains from prescriptive instructions. Qwen-32B remains stable, peaking at 24.78\% (\pStruct{Python}) with minimal variation across configurations. Qwen-7B favors \pReason{Python} (20.00\%), while other prompts decline to 16.09\%--16.96\%. Qwen-1.5B degrades from 16.96\% baseline to 12.17\% under \pEdge{Python}. Unlike Java, Python performance depends less on constraint density and more on whether prompts emphasize minimal operational requirements. Excessive guidance obscures rather than clarifies.


\begin{table*}[t]
\centering
\small
\renewcommand{\arraystretch}{1.2}
\setlength{\tabcolsep}{4pt}
\arrayrulecolor{black}
\resizebox{\textwidth}{!}{%
{\color{black}
\begin{tabular}{|l|c|ccc|ccc|ccc|ccc|}
\toprule
\rowcolor{headergray}
\textbf{Strategy}
& \multicolumn{13}{c|}{\textbf{Zero-shot Prompting}} \\
\rowcolor{headergray}

& \multicolumn{1}{c|}{\pname{Base}}
& \multicolumn{3}{c|}{\pname{Struct}}
& \multicolumn{3}{c|}{\pname{Robust}}
& \multicolumn{3}{c|}{\pname{Reason}}
& \multicolumn{3}{c|}{\pname{Edge}} \\
\rowcolor{headergray}
\textbf{Model}
& \multicolumn{1}{c|}{\textbf{Pass@1}}
& \textbf{Pass@1} & \textbf{$p$-value} & \textbf{OR}
& \textbf{Pass@1} & \textbf{$p$-value} & \textbf{OR}
& \textbf{Pass@1} & \textbf{$p$-value} & \textbf{OR}
& \textbf{Pass@1} & \textbf{$p$-value} & \textbf{OR} \\
\midrule

\multicolumn{14}{c}{\textbf{Java}} \\
\midrule


\mname{Qwen-1.5B}
& \bestJ{28.26}
& \worse{24.35} & \na & \na
& \worse{25.22} & \na & \na
& \worse{20.87} & \pv{$<0.05$} & \OR{2.2143}
& \worse{20.87} & \pv{$<0.05$} & \OR{2.5455}\\

\mname{Qwen-7B}
& 26.09
& \better{29.57} & \na & \na
& \better{\bestJ{32.61}} & \na & \na
& \better{30.00} & \na & \na
& \better{31.30} & \na & \na \\

\mname{Qwen-32B}
& 36.52
& \worse{34.78} & \na & \na
& \better{\bestJ{40.43}} & \na & \na
& \better{40.00} & \na & \na
& \better{38.70} & \na & \na \\

\midrule
\rowcolor{gptossrow}
\mname{\textbf{GPT-20B}}
& 37.83
& \better{41.30} & \na & \na
& \better{40.00} & \na & \na
& \better{39.57} & \na & \na
& \better{\bestJ{42.61}} & \na & \na \\

\midrule
\multicolumn{14}{c}{\textbf{Python}} \\
\midrule


\mname{Qwen-1.5B}
& \bestP{16.96}
& \worse{15.22} & \na & \na
& \worse{13.04} & \na & \na
& \worse{12.61} & \na & \na
& \worse{12.17} & \na & \na \\

\mname{Qwen-7B}
& 19.57
& \worse{16.96} & \na & \na
& \worse{16.09} & \na & \na
& \better{\bestP{20.00}} & \na & \na
& \worse{19.13} & \na & \na \\

\mname{Qwen-32B}
& 24.35
& \better{\bestP{24.78}} & \na & \na
& \worse{23.04} & \na & \na
& \worse{23.04} & \na & \na
& \worse{23.91} & \na & \na \\

\midrule
\rowcolor{gptossrow}
\mname{\textbf{GPT-20B}}
& \bestP{25.65}
& \worse{23.48} & \na & \na
& \worse{24.78} & \na & \na
& \worse{23.48} & \na & \na
& \worse{23.91} & \na & \na \\

\bottomrule
\end{tabular}
}}
\vspace{10pt}
\caption{\textbf{Performance results for the zero-shot prompting strategy, evaluated with \passat{1}{1}.}}
\label{tab:zero-pass@1-t1}
\end{table*}

\begin{table*}[t]
\centering
\small
\renewcommand{\arraystretch}{1.2}
\setlength{\tabcolsep}{4pt}
\arrayrulecolor{black}
\resizebox{\textwidth}{!}{%
{\color{black}
\begin{tabular}{|l|c|ccc|ccc|ccc|ccc|}
\toprule
\rowcolor{headergray}
\textbf{Strategy}
& \multicolumn{13}{c|}{\textbf{Fixed Prompting}} \\
\rowcolor{headergray}

& \multicolumn{1}{c|}{\pname{Base}}
& \multicolumn{3}{c|}{\pname{Struct}}
& \multicolumn{3}{c|}{\pname{Robust}}
& \multicolumn{3}{c|}{\pname{Reason}}
& \multicolumn{3}{c|}{\pname{Edge}} \\
\rowcolor{headergray}
\textbf{Model}
& \multicolumn{1}{c|}{\textbf{Pass@1}}
& \textbf{Pass@1} & \textbf{$p$-value} & \textbf{OR}
& \textbf{Pass@1} & \textbf{$p$-value} & \textbf{OR}
& \textbf{Pass@1} & \textbf{$p$-value} & \textbf{OR}
& \textbf{Pass@1} & \textbf{$p$-value} & \textbf{OR} \\
\midrule

\multicolumn{14}{c}{\textbf{Java}} \\
\midrule


\mname{Qwen-1.5B}
& 20.00
& \better{23.91} & \na & \na
& \better{22.61} & \na & \na
& \better{\bestJ{26.96}} & \na & \na
& \better{23.04} & \na & \na \\

\mname{Qwen-7B}
& 14.35
& \better{16.96} & \na & \na
& \better{18.26} & \na & \na
& \better{\bestJ{19.13}} & \na & \na
& \better{18.26} & \na & \na \\

\mname{Qwen-32B}
& 9.57
& \better{20.00} & \pv{$<0.05$} & \OR{0.2258}
& \better{22.61} & \pv{$<0.05$} & \OR{0.2683}
& \better{23.48} & \pv{$<0.05$} & \OR{0.1795}
& \better{\bestJ{24.35}} & \pv{$<0.05$} & \OR{0.2444} \\

\midrule
\rowcolor{gptossrow}
\mname{\textbf{GPT-20B}}
& 18.26
& \better{25.65} & \pv{$<0.05$} & \OR{0.5278}
& \better{30.87} & \pv{$<0.05$} & \OR{0.2368}
& \better{\bestJ{34.35}} & \pv{$<0.05$} & \OR{0.2128}
& \better{32.17} & \pv{$<0.05$} & \OR{0.2381} \\

\midrule
\multicolumn{14}{c}{\textbf{Python}} \\
\midrule


\mname{Qwen-1.5B}
& 13.48
& \worse{10.43} & \na & \na
& \worse{13.04}  & \na & \na
& \worse{11.74} & \na & \na
& \better{\bestP{15.65}}  & \na & \na \\

\mname{Qwen-7B}
& \bestP{20.43}
& \worse{19.13} & \na & \na
& \worse{18.70} & \na & \na
& \worse{16.96} & \na & \na
& \worse{19.57} & \na & \na \\

\mname{Qwen-32B}
& 23.04
& \worse{22.17} & \na & \na
& \worse{21.74} & \na & \na
& \better{\bestP{25.65}} & \na & \na
& \better{23.91} & \na & \na \\

\midrule
\rowcolor{gptossrow}
\mname{\textbf{GPT-20B}}
& 24.35
& \better{25.65} & \na & \na
& \better{\bestP{26.09}} & \na & \na
& \better{25.22} & \na & \na
& \better{25.22} & \na & \na \\

\bottomrule
\end{tabular}
}}
\vspace{10pt}
\caption{\textbf{Performance results for the 3-shot fixed prompting strategy, evaluated with \passat{1}{1}}.}
\label{tab:random-pass@1-t1}
\end{table*}

\begin{table*}[t]
\centering
\small
\renewcommand{\arraystretch}{1.2}
\setlength{\tabcolsep}{4pt}
\arrayrulecolor{black}
\resizebox{\textwidth}{!}{%
{\color{black}
\begin{tabular}{|l|c|ccc|ccc|ccc|ccc|}
\toprule
\rowcolor{headergray}
\textbf{Strategy}
& \multicolumn{13}{c|}{\textbf{Retrieval-based prompting}} \\
\rowcolor{headergray}

& \multicolumn{1}{c|}{\pname{Base}}
& \multicolumn{3}{c|}{\pname{Struct}}
& \multicolumn{3}{c|}{\pname{Robust}}
& \multicolumn{3}{c|}{\pname{Reason}}
& \multicolumn{3}{c|}{\pname{Edge}} \\
\rowcolor{headergray}
\textbf{Model}
& \multicolumn{1}{c|}{\textbf{Pass@1}}
& \textbf{Pass@1} & \textbf{$p$-value} & \textbf{OR}
& \textbf{Pass@1} & \textbf{$p$-value} & \textbf{OR}
& \textbf{Pass@1} & \textbf{$p$-value} & \textbf{OR}
& \textbf{Pass@1} & \textbf{$p$-value} & \textbf{OR} \\
\midrule

\multicolumn{14}{c}{\textbf{Java}} \\
\midrule

\mname{Qwen-1.5B}
& 18.70
& \better{\bestJ{24.35}} & \na & \na
& \better{22.61} & \na & \na
& \better{20.43} & \na & \na
& \better{22.61} & \na & \na \\

\mname{Qwen-7B}
& 9.57
& \better{14.35} & \na & \na
& \better{14.78} & \na & \na
& \better{16.52} & \pv{$<0.05$} & \OR{0.4074}
& \better{\bestJ{18.26}} & \pv{$<0.05$} & \OR{0.3103} \\

\mname{Qwen-32B}
& 16.96
& \better{26.52} & \pv{$<0.05$} & \OR{0.3529}
& \better{\bestJ{33.48}} & \pv{$<0.05$} & \OR{0.1739}
& \better{31.30} & \pv{$<0.05$} & \OR{0.1951}
& \better{32.61} & \pv{$<0.05$} & \OR{0.1429} \\

\midrule
\rowcolor{gptossrow}
\mname{\textbf{GPT-20B}}
& 27.83
& \better{35.22} & \pv{$<0.05$} & \OR{0.4333}
& \better{36.52} & \pv{$<0.05$} & \OR{0.3548}
& \better{41.30} & \pv{$<0.05$} & \OR{0.2051}
& \better{\bestJ{44.78}} & \pv{$<0.05$} & \OR{0.1333} \\

\midrule
\multicolumn{14}{c}{\textbf{Python}} \\
\midrule

\mname{Qwen-1.5B}
& \bestP{15.22}
& \worse{14.78} & \na & \na
& \worse{14.35} & \na & \na
& \worse{12.17} & \na & \na
& \worse{12.17} & \na & \na \\

\mname{Qwen-7B}
& 16.52
& \better{16.96} & \na & \na
& 16.52 -- & \na & \na
& \better{\bestP{20.00}} & \na & \na
& \better{19.13} & \na & \na \\

\mname{Qwen-32B}
& 23.48
& \better{\bestP{24.78}} & \na & \na
& \better{24.35} & \na & \na
& \worse{22.61} & \na & \na
& \better{\bestP{24.78}} & \na & \na \\

\midrule
\rowcolor{gptossrow}
\mname{\textbf{GPT-20B}}
& 21.30
& \better{\bestP{23.04}} & \na & \na
& \better{21.74} & \na & \na
& \worse{20.87} & \na & \na
& \better{\bestP{23.04}} & \na & \na \\

\bottomrule
\end{tabular}
}}
\vspace{10pt}
\caption{\textbf{Performance results for the 3-shot retrieval-based prompting strategy, evaluated with \passat{1}{1}.}}
\label{tab:retrieval-pass@1-t1}
\end{table*}

Fixed 3-shot examples expose sharp, language-dependent behaviors (\tabref{tab:random-pass@1-t1}). In Java, every backbone benefits from stronger system guidance, but the \pBase{Java} condition is strikingly brittle for larger models. Qwen-32B, for instance, delivers only 9.57\% Pass@1 with \pBase{Java} -- roughly one correct solution out of ten -- yet more informative prompts steadily pull it back, reaching 24.35\% with \pEdge{Java}. Statistical tests confirm this improvement is reliable: all enriched prompts show significant gains over \pBase{Java}, with odds ratios indicating a $3.7-5.6\times$ higher likelihood of generating correct solutions. Qwen-7B follows the same direction (14.35\% to 19.13\%, best with \pReason{Java}), while Qwen-1.5B peaks higher at 26.96\% under \pReason{Java} but without statistically significant differences in this fixed-example setting. 

GPT-20B shows the clearest ``re-anchoring'' effect: it climbs from 18.26\% with \pBase{Java} to a best of 34.35\% with \pReason{Java}, with all richer prompts significant ($p<0.05$) and ORs ranging from 1.89 up to 4.7. Overall, with fixed examples and $T=1$, Java correctness depends strongly on whether the system prompt can filter exemplar-induced noise; minimal prompting leaves the larger backbones under-anchored, whereas richer prompts restore functional alignment in a statistically reliable way.

Python reveals a contrasting pattern in the same table, with smaller and inconsistent gains that fail to reach statistical significance. Qwen-32B shows modest directional improvement from reasoning guidance (23.04\% to 25.65\%), but the effect remains statistically inconclusive. Qwen-7B actually performs best with minimal prompting (20.43\% at \pBase{Python}), declining to 16.96\% under \pReason{Python} as prescriptive constraints interfere with performance. Qwen-1.5B responds selectively, improving only under \pEdge{Python} (15.65\% vs. 13.48\% baseline). Even GPT-20B, while relatively stable across prompts (peaking at 26.09\% with \pRobust{Python}), shows no statistically significant differences.

The pattern suggests that for Python, fixed examples already capture sufficient structural information. Additional constraints compete with rather than complement this signal, particularly for mid-sized models. Only the largest model (Qwen-32B) shows any directional benefit from reasoning-oriented prompts, though even this remains statistically uncertain.

Switching to retrieval-based examples (\tabref{tab:retrieval-pass@1-t1}) transforms performance patterns in Java. Qwen-32B climbs from 16.96\% (\pBase{Java}) to 33.48\% (\pRobust{Java}), while GPT-20B rises from 27.83\% to 44.78\% (\pEdge{Java})---both with statistically significant gains (p<0.05) when retrieval pairs with descriptive prompts.

Smaller models benefit less. Qwen-1.5B improves modestly to 24.35\% (\pStruct{Java}), while Qwen-7B reaches only 18.26\% despite relevant examples. This reveals a capacity threshold: retrieval helps only when models can extract meaningful signals from retrieved context.

In Python, retrieval does not provide a universal boost, and the statistical evidence remains uniformly non-significant. Qwen-32B stays stable and again favors simple structure (24.78\% with \pStruct{Python} and 24.78\% with \pEdge{Python}), while \pReason{Python} drops to 22.61\%. Qwen-7B retains its preference for \pReason{Python} (20.00\%), but the baseline \pBase{Python} falls to 16.52\%, suggesting that retrieved examples can shift reliance away from minimal prompting without guaranteeing higher accuracy. Qwen-1.5B continues to degrade with more information-dense prompts, reaching 12.17\% under \pReason{Python} and \pEdge{Python}. GPT-20B shows only limited movement in Python (best at 23.04\% with \pStruct{Python} and \pEdge{Python}), consistent with the overall compression observed in this language. 

\begin{tcolorbox}[
  enhanced,
  breakable,
  colback=cyan!8,        
  colframe=black,        
  coltext=black,
  arc=8pt,               
  boxrule=0.8pt,         
  left=6pt,right=6pt,top=10pt,bottom=8pt,
  fonttitle=\bfseries,
  title=Answer to RQ2.2,   
  coltitle=black,
  varwidth boxed title,  
  attach boxed title to top left={yshift=-3mm,xshift=6mm},
    boxed title style={
        colback=gray!40,
        boxrule=0.7pt,
        arc=8pt,
        outer arc=8pt,
        left=5pt,
        right=5pt,
        top=0.5pt,
        bottom=0.5pt,
    }
]

These results paint a coherent picture of how prompting strategy, example selection, and model scale interact under stochastic generation. Overall, at $T=1$, examples (\ie shots) and prompt specificity interact most strongly in Java -- with McNemar's tests confirming large, statistically reliable improvements over \pBase{} -- whereas Python code remains comparatively sound, showing smaller and statistically inconclusive changes. The language-dependent sensitivity suggests that Java's more verbose syntax and stricter type requirements amplify the impact of both example quality and prompt guidance.
\end{tcolorbox}
\subsubsection{\textbf{RQ2.3: \passat{5}{1}}}

\tabref{tab:zero-pass@5-t1} presents zero-shot Pass@5 results, where the interplay of sampling and prompt sensitivity reveals whether prompts help models explore productive solution variations or constrain all attempts toward similar failures. Java shows the strongest prompt leverage, particularly for larger models. Qwen-32B climbs from 42.61\% (\pBase{Java}) to 47.39\% (\pRobust{Java}), with \pReason{Java} (46.96\%) and \pEdge{Java} (46.52\%) maintaining similarly high performance. The improvement under \pRobust{Java} is statistically significant, indicating roughly $4.7\times$  higher odds of generating at least one passing solution compared to baseline.

Qwen-7B follows a similar pattern, improving from 41.30\% to 44.35\% (\pRobust{Java}), with \pStruct{Java} at 43.91\%, though these gains lack statistical significance. Notably, \pEdge{Java} provides no improvement (41.30\%), indicating that overly specific constraints offer diminishing returns under sampling. Qwen-1.5B reveals the opposite behavior: it peaks at 42.61\% (\pStruct{Java}) but degrades as prompts become more information-dense -- dropping to 36.09\% (\pRobust{Java}), 37.83\% (\pReason{Java}), and 36.96\% (\pEdge{Java}). For the smallest model, structural framing is optimal -- additional constraints excessively narrow the solution space, causing generation attempts to collapse on the same incorrect patterns rather than exploring simpler plausible correct solutions.

For Python code generation, prompt variations have minimal impact, with most models showing similar performance across different prompts. The sole statistically significant exception occurs with Qwen-1.5B, the smallest model in the family, which clearly performs well in a less ``information-dense'' setting (\pEdge{Python} vs. \pBase{Python}).


\begin{table*}[t]
\centering
\small
\renewcommand{\arraystretch}{1.2}
\setlength{\tabcolsep}{4pt}
\arrayrulecolor{black}
\resizebox{\textwidth}{!}{%
{\color{black}
\begin{tabular}{|l|c|ccc|ccc|ccc|ccc|}
\toprule
\rowcolor{headergray}
\textbf{Strategy}
& \multicolumn{13}{c|}{\textbf{Zero-shot Prompting}} \\
\rowcolor{headergray}

& \multicolumn{1}{c|}{\pname{Base}}
& \multicolumn{3}{c|}{\pname{Struct}}
& \multicolumn{3}{c|}{\pname{Robust}}
& \multicolumn{3}{c|}{\pname{Reason}}
& \multicolumn{3}{c|}{\pname{Edge}} \\
\rowcolor{headergray}
\textbf{Model}
& \multicolumn{1}{c|}{\textbf{Pass@5}}
& \textbf{Pass@5} & \textbf{$p$-value} & \textbf{OR}
& \textbf{Pass@5} & \textbf{$p$-value} & \textbf{OR}
& \textbf{Pass@5} & \textbf{$p$-value} & \textbf{OR}
& \textbf{Pass@5} & \textbf{$p$-value} & \textbf{OR} \\
\midrule

\multicolumn{14}{c}{\textbf{Java}} \\
\midrule


\mname{Qwen-1.5B}
& 40.43
& \better{\bestJ{42.61}} & \na & \na
& \worse{36.09} & \na & \na
& \worse{37.83} & \na & \na
& \worse{36.96} & \na & \na\\

\mname{Qwen-7B}
& 41.30
& \better{43.91} & \na & \na
& \better{\bestJ{44.35}} & \na & \na
& \better{42.61} & \na & \na
& 41.30 -- & \na & \na \\

\mname{Qwen-32B}
& 42.61
& \better{45.65} & \na & \na
& \better{\bestJ{47.39}} & \pv{$<0.05$} & \OR{0.2143}
& \better{46.96} & \na & \na
& \better{46.52} & \na & \na \\
\midrule

\midrule
\rowcolor{gptossrow}
\mname{\textbf{GPT-20B}}
& 48.70
& \better{49.57} & \na & \na
& \better{\bestJ{50.43}} & \na & \na
& \better{49.57} & \na & \na
& 48.70 -- & \na & \na \\

\midrule
\multicolumn{14}{c}{\textbf{Python}} \\
\midrule

\mname{Qwen-1.5B}
& 26.96
& \better{\bestP{27.83}} & \na & \na
& \worse{25.22} & \na & \na
& \worse{25.22} & \na & \na
& \worse{20.87} & \pv{$<0.05$} & \OR{3.00} \\

\mname{Qwen-7B}
& 31.74
& \better{\bestP{32.17}} & \na & \na
& \worse{31.30} & \na & \na
& \better{\bestP{32.17}} & \na & \na
& \better{\bestP{32.17}} & \na & \na \\

\mname{Qwen-32B}
& 31.30
& \better{\bestP{32.17}} & \na & \na
& 31.30 -- & \na & \na
& \better{\bestP{32.17}} & \na & \na
& \better{31.74} & \na & \na \\

\midrule
\rowcolor{gptossrow}
\mname{\textbf{GPT-20B}}
& 33.91
& \better{34.35} & \na & \na
& \better{\bestP{34.78}} & \na & \na
& \better{34.35} & \na & \na
& 33.91 -- & \na & \na \\

\bottomrule
\end{tabular}
}}
\vspace{10pt}
\caption{\textbf{Performance results for the zero-shot prompting strategy, evaluated with \passat{5}{1}.}}
\label{tab:zero-pass@5-t1}
\end{table*}

\begin{table*}[t]
\centering
\small
\renewcommand{\arraystretch}{1.2}
\setlength{\tabcolsep}{4pt}
\arrayrulecolor{black}
\resizebox{\textwidth}{!}{%
{\color{black}
\begin{tabular}{|l|c|ccc|ccc|ccc|ccc|}
\toprule
\rowcolor{headergray}
\textbf{Strategy}
& \multicolumn{13}{c|}{\textbf{Fixed Prompting}} \\
\rowcolor{headergray}

& \multicolumn{1}{c|}{\pname{Base}}
& \multicolumn{3}{c|}{\pname{Struct}}
& \multicolumn{3}{c|}{\pname{Robust}}
& \multicolumn{3}{c|}{\pname{Reason}}
& \multicolumn{3}{c|}{\pname{Edge}} \\
\rowcolor{headergray}
\textbf{Model}
& \multicolumn{1}{c|}{\textbf{Pass@5}}
& \textbf{Pass@5} & \textbf{$p$-value} & \textbf{OR}
& \textbf{Pass@5} & \textbf{$p$-value} & \textbf{OR}
& \textbf{Pass@5} & \textbf{$p$-value} & \textbf{OR}
& \textbf{Pass@5} & \textbf{$p$-value} & \textbf{OR} \\
\midrule

\multicolumn{14}{c}{\textbf{Java}} \\
\midrule


\mname{Qwen-1.5B}
& 35.22
& \better{\bestJ{42.61}} & \pv{$<0.05$} & \OR{0.2273}
& \better{40.87} & \pv{$<0.05$} & \OR{0.2353}
& \better{37.83} & \na & \na
& \better{38.70} & \na & \na \\

\mname{Qwen-7B}
& 31.74
& \better{36.52} & \na & \na
& \better{33.91} & \na & \na
& \better{36.52} & \na & \na
& \better{\bestJ{38.70}} & \pv{$<0.05$} & \OR{0.3846} \\

\mname{Qwen-32B}
& 22.17
& \better{33.04} & \pv{$<0.05$} & \OR{0.2188}
& \better{\bestJ{42.17}} & \pv{$<0.05$} & \OR{0.098}
& \better{39.13} & \pv{$<0.05$} & \OR{0.1333}
& \better{41.74} & \pv{$<0.05$} & \OR{0.10} \\
\midrule

\midrule
\rowcolor{gptossrow}
\mname{\textbf{GPT-20B}}
& 45.65
& \better{46.52} & \na & \na
& \better{\bestJ{50.43}} & \na & \na
& \better{47.39} & \na & \na
& \better{47.57} & \na & \na \\

\midrule
\multicolumn{14}{c}{\textbf{Python}} \\
\midrule

\mname{Qwen-1.5B}
& 21.74
& \better{22.17} & \na & \na
& \better{\bestP{28.26}}  & \pv{$<0.05$} & \OR{0.2105}
& \better{26.52} & \na & \na
& \better{26.09}  & \na & \na \\

\mname{Qwen-7B}
& 30.00
& \better{\bestP{32.17}} & \na & \na
& \worse{29.57} & \na & \na
& \better{30.43} & \na & \na
& \better{30.43} & \na & \na \\

\mname{Qwen-32B}
& 30.43
& \worse{30.00} & \na & \na
& \better{\bestP{31.30}} & \na & \na
& \better{30.87} & \na & \na
& \worse{30.00} & \na & \na \\

\midrule
\rowcolor{gptossrow}
\mname{\textbf{GPT-20B}}
& \bestP{36.09}
& \worse{33.04} & \na & \na
& \worse{34.78} & \na & \na
& \worse{35.65} & \na & \na
& \worse{33.04} & \na & \na \\

\bottomrule
\end{tabular}
}}
\vspace{10pt}
\caption{\textbf{Performance results for the 3-shot fixed prompting strategy, evaluated with \passat{5}{1}.}} 
\label{tab:random-pass@5-t1}
\end{table*}

Introducing fixed examples (\tabref{tab:random-pass@5-t1}) reveals how examples and prompts interact under sampling. In Java, Qwen-32B shows the strongest sensitivity: performance drops to 22.17\% with \pBase{Java}, but stronger prompts recover dramatically -- reaching 42.17\% (\pRobust{Java}, OR = 10.20, $p<0.05$) and 41.74\% (\pEdge{Java}, OR = 10.00, $p<0.05$). These large ORs indicate the model is approximately $10 times$ more likely to generate at least one passing solution with robust-handling or edge-coverage prompts. This suggests fixed examples can inject misleading priors that persist across sampled outputs unless prompts strongly reassert constraints. \pStruct{Java} shows weaker recovery at 33.04\% (OR = 4.57, $p<0.05$).

Qwen-7B benefits selectively, with significant improvement only under \pEdge{Java} (38.70\%, OR = 2.60, $p<0.05$), while other enriched prompts improve directionally without statistical significance. Qwen-1.5B demonstrates a different pattern, benefiting from both \pStruct{Java} (42.61\%, OR = 4.40, $p<0.05$) and \pRobust{Java} (40.87\%, OR = 4.25, $p<0.05$). For the smallest model, fixed examples provide helpful scaffolding when prompts maintain interface alignment.

Python shows minimal prompt sensitivity under fixed examples. Only Qwen-1.5B exhibits significant improvement (\pRobust{Python}: 28.26\% vs. 21.74\%, OR = 4.75, $p<0.05$), while Qwen-7B and Qwen-32B show no statistical differences across prompts, clustering around 30--32\%. Even GPT-20B degrades in Python (36.09\% to 33.04\%), suggesting fixed examples are less effective in Python regardless of model capacity.


\begin{table*}[t]
\centering
\small
\renewcommand{\arraystretch}{1.2}
\setlength{\tabcolsep}{4pt}
\arrayrulecolor{black}
\resizebox{\textwidth}{!}{%
{\color{black}
\begin{tabular}{|l|c|ccc|ccc|ccc|ccc|}
\toprule
\rowcolor{headergray}
\textbf{Strategy}
& \multicolumn{13}{c|}{\textbf{Retrieval-based prompting}} \\
\rowcolor{headergray}

& \multicolumn{1}{c|}{\pname{Base}}
& \multicolumn{3}{c|}{\pname{Struct}}
& \multicolumn{3}{c|}{\pname{Robust}}
& \multicolumn{3}{c|}{\pname{Reason}}
& \multicolumn{3}{c|}{\pname{Edge}} \\
\rowcolor{headergray}
\textbf{Model}
& \multicolumn{1}{c|}{\textbf{Pass@5}}
& \textbf{Pass@5} & \textbf{$p$-value} & \textbf{OR}
& \textbf{Pass@5} & \textbf{$p$-value} & \textbf{OR}
& \textbf{Pass@5} & \textbf{$p$-value} & \textbf{OR}
& \textbf{Pass@5} & \textbf{$p$-value} & \textbf{OR} \\
\midrule

\multicolumn{14}{c}{\textbf{Java}} \\
\midrule

\mname{Qwen-1.5B}
& 38.70
& \better{\bestJ{42.17}} & \na & \na
& 38.70 -- & \na & \na
& \better{39.60} & \na & \na
& \better{40.00} & \na & \na \\

\mname{Qwen-7B}
& 24.78
& \better{32.17} & \pv{$<0.05$} & \OR{0.3704}
& \better{\bestJ{34.35}} & \pv{$<0.05$} & \OR{0.2903}
& \better{29.57} & \na & \na
& \better{\bestJ{34.35}} & \pv{$<0.05$} & \OR{0.3592} \\

\mname{Qwen-32B}
& 31.30
& \better{40.87} & \pv{$<0.05$} & \OR{0.2667}
& \better{45.65} & \pv{$<0.05$} & \OR{0.1081}
& \better{\bestJ{46.96}} & \pv{$<0.05$} & \OR{0.0526}
& \better{46.09} & \pv{$<0.05$} & \OR{0.1282} \\
\midrule

\midrule
\rowcolor{gptossrow}
\mname{\textbf{GPT-20B}}
& 46.52
& \better{47.83} & \na & \na
& \better{50.00} & \na & \na
& \better{50.87} & \na & \na
& \better{\bestJ{53.91}} & \pv{$<0.05$} & \OR{0.1053} \\

\midrule
\multicolumn{14}{c}{\textbf{Python}} \\
\midrule

\mname{Qwen-1.5B}
& \bestP{26.52}
& \worse{25.65} & \na & \na
& \worse{25.65} & \na & \na
& \worse{24.35} & \na & \na
& \worse{23.91} & \na & \na \\

\mname{Qwen-7B}
& 30.43
& \better{\bestP{33.04}} & \na & \na
& \better{31.74} & \na & \na
& \better{31.74} & \na & \na
& \better{30.87} & \na & \na \\

\mname{Qwen-32B}
& 30.87
& \better{31.74} & \na & \na
& \better{32.61} & \na & \na
& \better{\bestP{33.04}} & \na & \na
& \better{31.30}  & \na & \na \\

\midrule
\rowcolor{gptossrow}
\mname{\textbf{GPT-20B}}
& 33.91
& \better{35.65} & \na & \na
& \better{34.78} & \na & \na
& \better{\bestP{36.52}} & \na & \na
& \better{34.35} & \na & \na \\

\bottomrule
\end{tabular}
}}
\vspace{10pt}
\caption{\textbf{Performance results for the 3-shot retrieval-based prompting strategy, evaluated with \passat{5}{1}.}}
\label{tab:retrieval-pass@5-t1}
\end{table*}

\tabref{tab:retrieval-pass@5-t1} presents results for retrieval-based prompting under Pass@5, where examples are selected based on semantic similarity to each query and the model generates five candidate solutions per problem. This regime tests whether query-relevant context combined with multiple sampling attempts can amplify prompt effectiveness.

In Java, retrieval-based prompting shows clear benefits for larger models, with statistically significant improvements across most enriched prompts. Qwen-32B demonstrates the strongest effect: code generated with \pReason{Java} is $19 \times$ more likely to be functionally correct compared to the same model using only \pBase{Java}.
Qwen-7B shows significant gains under \pStruct{Java} (32.17\%, OR = 2.70, $p<0.05$), \pRobust{Java} (34.35\%, OR = 3.44, $p<0.05$), and \pEdge{Java} (34.35\%, OR = 2.78, $p<0.05$), though improvements remain more modest than the largest open-source model. Qwen-1.5B improves to 42.17\% (\pStruct{Java}) without statistical significance, confirming the smallest model's limited capacity to reliably exploit retrieved context under sampling, despite instruction-based specialization.

On the other hand, we find that GPT-20B -- despite the absence of code-centric specialization -- demonstrates that learning how to translate natural language intents into code extends beyond code-specialized models. In this regard, GPT-20B shows similar sensitivity to enriched prompts. Starting from 46.52\%, it reaches 53.91\% (\pEdge{Java}, OR = 9.50, $p<0.05$) -- the only statistically significant result among its enriched prompts. The fact that GPT-20B requires the most constraint-rich prompt to achieve significance under Pass@5, while showing gains across all enriched variants, suggests that larger models need more prescriptive constraints to ensure consistent improvements across multiple sampling attempts rather than sporadic gains. This indicates that the threshold for statistical reliability differs across model scales: larger models may occasionally produce better outputs with lighter prompts, but require stronger constraints to reliably improve the majority of their five generated candidates.

Looking at the results achieved in the context of Python-based code generation,  we found no statistically significant improvements for any model, with performance remaining tightly clustered. Among the Qwen family, both Qwen-7B and Qwen-32B reach similar peaks (33.04\% each), though via different prompts -- Qwen-7B achieves its best with \pStruct{Python}, while Qwen-32B peaks under \pReason{Python}. Qwen-1.5B progressively degrades from its 26.52\% baseline to 23.91\% (\pEdge{Python}) as constraints described in system prompts increase.

GPT-20B exhibits the same resistance to prompt-based guidance in Python observed across the Qwen family. Despite its superior baseline (33.91\%) and peak performance (36.52\% under \pReason{Python}) -- reflecting its strong general capabilities even as a non-code-specialized model -- GPT-20B shows no statistically significant differences across any prompt variants. This consistency across model families, from the smallest Qwen-1.5B to state-of-the-art GPT-20B, confirms that Python code generation remains largely insensitive to prompt specificity regardless of model scale or specialization. The pattern suggests that when solving the same algorithmic challenges, models respond fundamentally differently depending on the target language: Java's verbose syntax and explicit type requirements create multiple intervention points where prompts can guide generation toward correctness, while Python's compact, implicit design offers fewer such leverage points, limiting the effectiveness of constraint-based steering even when examples are query-relevant and multiple candidates are sampled.

\begin{tcolorbox}[
  enhanced,
  breakable,
  colback=cyan!8,        
  colframe=black,        
  coltext=black,
  arc=8pt,               
  boxrule=0.8pt,         
  left=6pt,right=6pt,top=10pt,bottom=8pt,
  fonttitle=\bfseries,
  title=Answer to RQ2.3,   
  coltitle=black,
  varwidth boxed title,  
  attach boxed title to top left={yshift=-3mm,xshift=6mm},
    boxed title style={
        colback=gray!40,
        boxrule=0.7pt,
        arc=8pt,
        outer arc=8pt,
        left=5pt,
        right=5pt,
        top=0.5pt,
        bottom=0.5pt,
    }
]
Pass@5 at $T=1$ confirms the Pass@1 story, with larger effect sizes under sampling. Java benefits substantially—especially with retrieval-based prompting and descriptive system prompts (up to 19$\times$ higher odds for Qwen-32B; GPT-20B reaches 53.91\%, which translates into the ability to generate correct code with $\sim 9\times$ higher odds than the one achieved by the same model operating with basic instruction (\pBase{}). Fixed examples are brittle unless paired with robustness constraints. Python shows no statistically significant gains in any setting, reinforcing that language structure (Java’s verbosity vs.\ Python’s terseness) primarily determines prompt leverage.\end{tcolorbox}

\begin{figure}[htbp]
    \centering
    \includegraphics[width=0.66\columnwidth]{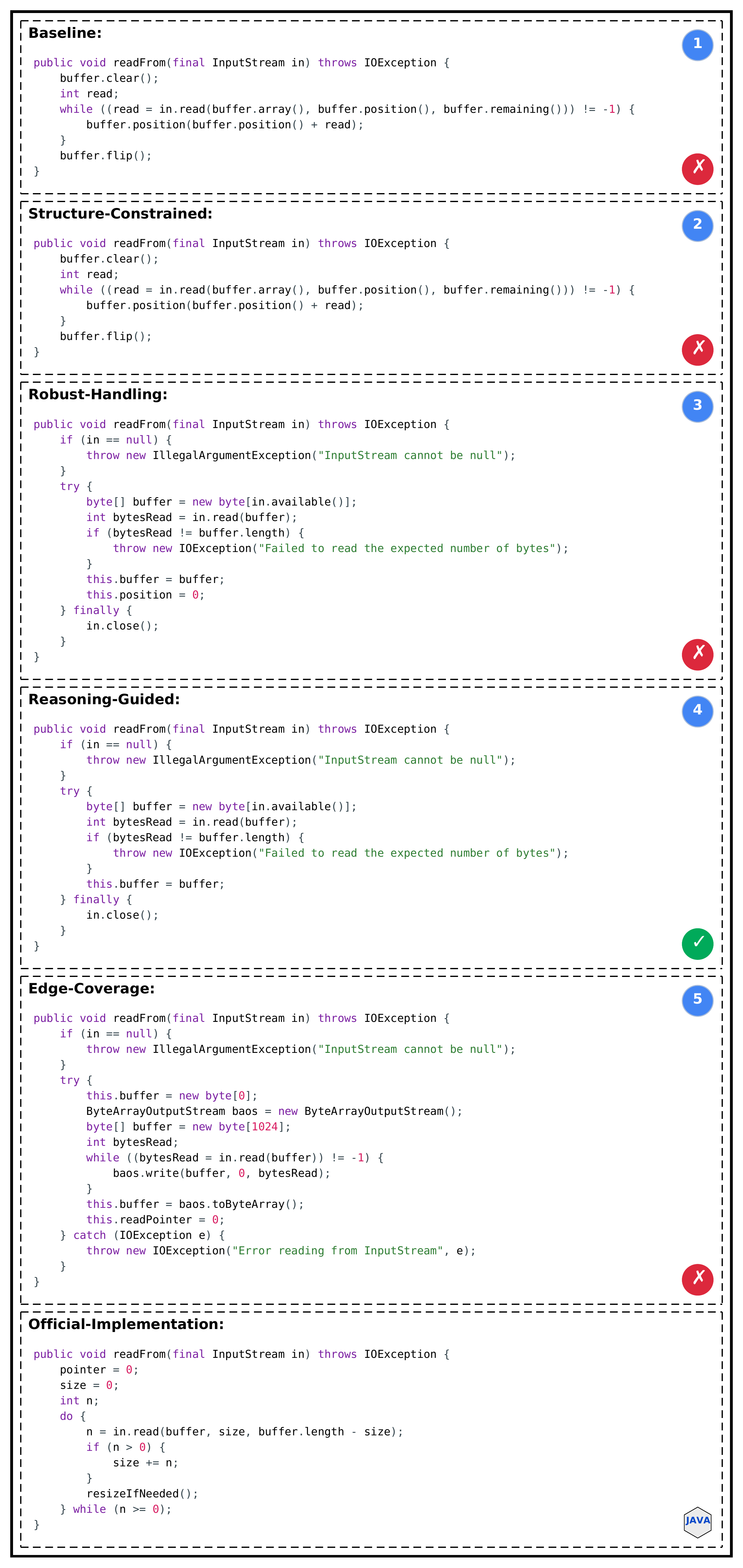}
    \caption{Example of generated Java code under varying system prompts.}
    \label{fig:example}
\end{figure}
\section{Qualitative Analysis}\label{sec:overll}
In this section, we complement the quantitative findings of Section~\ref{sec:results} with a qualitative analysis of a representative failure case (\figref{fig:example}). While our aggregate results demonstrate that prompt choice significantly affects pass rates across the benchmark, this case study reveals the mechanisms underlying such variation. To illustrate these mechanisms, we selected a Java instance from the \textit{zero-shot} setting that exhibits maximal prompt sensitivity: a problem where the strongest model in our study, Qwen-32B, produces a test-passing solution under exactly one of five system prompts, while each of the remaining four prompts induces a distinct, traceable failure mode. 

We note that -- consistent with our goal of isolating prompt-induced effects -- we anchor the discussion on the simplest and least prescriptive prompt (\pBase{}) as a reference point, using it to highlight how progressively enriched prompts reshape the model's decisions and introduce specific failure modes. 


In \figref{fig:example} -- the five generated variants correspond to the five system prompts in the order \texttt{Base}, \texttt{Struct}, \texttt{Robust}, \texttt{Reason}, and \texttt{Edge}. Only the \texttt{Reason} variant passed all tests. Below, we analyze why the other remaining four are incorrect, and why the \texttt{Reason} implementation is uniquely aligned with the unit-test expectations despite syntactically deviating from the official reference--a problem originally investigated by Mastropaolo \etal \cite{mastropaolo2023robustness}.

We now examine these findings in detail. Consider Example \stepbadge{1} depicted in \figref{fig:example}, which shows the code generated by Qwen-32B under the \texttt{Base} system prompt. Comparing this output against the ground-truth implementation (bottom of the figure) reveals a fundamental architectural mismatch. The model's solution treats \texttt{buffer} as a \texttt{ByteBuffer} object and invokes standard buffer manipulation methods (\texttt{clear}, \texttt{array}, \texttt{position}, \texttt{remaining}, \texttt{flip}) to manage reads and state transitions. However, the official implementation defines \texttt{buffer} as a primitive byte array that is dynamically grown and populated via \texttt{resizeIfNeeded}.

This discrepancy has cascading consequences. To begin with, if the class follows the official field layout, the generated code will not compile because \texttt{ByteBuffer} methods cannot be invoked on a byte array. Second, even assuming a hypothetical compatible \texttt{ByteBuffer} field exists, the implementation is logically incorrect: it never maintains the required \texttt{pointer} and \texttt{size} fields that track the current write position and occupied capacity. Third, the solution exhibits a critical capacity assumption, it presumes a fixed-size buffer can accommodate the entire input stream, completely omitting the resizing mechanism. This is not a stylistic choice but a specification violation: the ground-truth implementation includes \texttt{resizeIfNeeded} precisely because stream length is unknown at read time, requiring dynamic memory allocation. Finally, the generated code performs no input validation (\eg null checks) and leaves resource management ambiguous—it does not clearly establish whether stream closure is the method's responsibility or the caller's, both common sources of test failures.

The \texttt{Struct} output--(Ex. \stepbadge{2} in \figref{fig:example})--is identical to \texttt{Base} in this case and therefore fails for the same reasons. Qualitatively, this is instructive: \textit{even when the prompt emphasizes structural compliance, the generated code still departs from the reference at the level of data representation and state invariants.} The method remains incompatible with a byte-array-backed implementation, does not reset or track the official state variables, and does not provide any growth path that would make it robust to larger inputs.

When prompted with \texttt{Robust}--(Ex. \stepbadge{3} in \figref{fig:example})--the model produces a substantially improved implementation compared to its \texttt{Base} output. The additional constraints communicated by \texttt{Robust} guide the model toward a byte-array architecture: it allocates buffer space using \texttt{in.available()}, performs a single \texttt{read} operation, validates that the bytes read match the allocated length, assigns the result to \texttt{this.buffer}, and ensures stream closure within a \texttt{finally} block for proper resource management. Yet, it does not pass all test cases, and the reasons lie in the statement \texttt{this.position = 0;}.
To understand why this single line is decisive, we leverage a key empirical constraint: the \texttt{Reason} and \texttt{Robust} implementations are otherwise identical, yet only \texttt{Reason} passes. This isolates the failure mechanism to the role of \texttt{position} in the surrounding class invariants and in the tests that follow \texttt{readFrom}.  Relative to the official solution, it attempts to reset \texttt{position}, which is not part of the official reference state. If the class is designed around \texttt{pointer} and \texttt{size}, resetting an unrelated field is either a compilation failure if the field does not exist, or a semantic failure if it exists but does not control the read cursor used by downstream methods.

The solution generated under \pEdge{Java}--(Ex. \stepbadge{5} in \figref{fig:example})-- demonstrates improved task comprehension through its use of a \texttt{ByteArrayOutputStream} loop that reads until end-of-stream. This design choice represents a clear advancement over \texttt{Robust}: it eliminates the brittle dependency on \texttt{available()} and correctly aggregates bytes across multiple read operations, aligning more closely with streaming semantics. Despite this architectural progress, the implementation still violates the specification in ways that expose fundamental misalignment with the reference design. First, it does not align with the reference state management: instead of resetting and updating \texttt{pointer} and \texttt{size}, it resets a different cursor field, \texttt{readPointer}. If the evaluation class skeleton does not define \texttt{readPointer}, this again yields an immediate compile-time failure. If \texttt{readPointer} exists, it is still not evidence that it is the authoritative cursor used by the rest of the class. Second, it does not close the input stream, leaving ownership ambiguous; this may be tolerated in some settings, but it can also be penalized in unit tests that explicitly check resource handling.

As anticipated, the only test-passing implementation is the \texttt{Reason} variant--(Ex. \stepbadge{4} in \figref{fig:example})--which reads \texttt{in.available()} bytes, performs a single \texttt{read}, checks for exact length, assigns \texttt{this.buffer}, and closes the stream in a \texttt{finally} block. This solution is not equivalent to the official answer in a general \texttt{InputStream} sense, since it relies on \texttt{available()} and assumes a full-length read. Its success suggests that the unit tests validate only a narrower set of behaviors than those addressed by the reference solution. A consistent explanation is that the harness supplies an \texttt{InputStream} whose \texttt{available()} accurately reflects the remaining content and whose \texttt{read} typically fills the requested array, such as an in-memory stream or a deterministic wrapper. Under such a harness, the \texttt{Reason} implementation achieves two properties that the failing variants do not simultaneously satisfy. First, it guarantees stream closure, which is absent in the \texttt{Base} and \texttt{Struct} outputs. Second, it avoids introducing potentially invalid state writes to non-reference fields such as \texttt{position} or \texttt{readPointer}. In a setting where the state variables are strictly named and the tests primarily validate buffer content and resource handling, refraining from mutating mismatched cursor fields can be safer than attempting to reset the wrong variable.

This example illustrates a sharp qualitative separation between \emph{reference-correct} implementations and \emph{test-aligned} implementations. The official answer embodies a fully general streaming design by repeatedly reading, resizing as needed, and maintaining explicit cursor invariants through \texttt{pointer} and \texttt{size}. The four failing variants each violate at least one essential requirement, most commonly by using an incompatible buffer representation, by writing to non-existent or non-authoritative state fields, or by deviating from harness-sensitive behaviors such as exception and resource handling. The \texttt{Reason} variant passes because it aligns with the most plausible evaluation constraints in this setting, particularly deterministic stream behavior compatible with \texttt{available()} and the explicit stream-closure behavior, even though the same strategy would be brittle outside this constrained test environment.

\section{Threats to Validity} \label{sec:threats}

\textbf{Construct validity} concerns whether our system prompts accurately represent different levels of instruction specificity. The five prompts were designed through manual judgment to reflect varying degrees of constraint detail. While performance differences suggest these variations had measurable effects, we cannot fully confirm how models internally interpret these distinctions. 

\textbf{Internal validity} relates to confounding factors within our experimental design. 
Different retrievers or larger corpora might yield different results. Additionally, the evaluated LLMs vary in architecture, tokenization, and training data, which may introduce model-specific sensitivities beyond system prompt design alone. We controlled for this by maintaining identical experimental parameters--temperature, sampling seeds, and evaluation metrics--across all configurations.

\textbf{Conclusion validity} concerns the reliability of our statistical inferences. Our hypothesis testing relies on McNemar's test over paired binary outcomes and reports matched-pairs odds ratios with confidence intervals, with Holm correction applied for multiple comparisons. This strengthens control of false positives, but it can reduce statistical power, increasing the chance of failing to detect modest but real effects, especially when effects are heterogeneous across languages, prompting strategies, and model scales.


\textbf{External validity} concerns generalizability beyond our experimental scope. Our findings focus on code generation in Java and Python using five specific system prompts and three prompting strategies. Results may not transfer to other languages (\eg C++, JavaScript) or tasks (\eg code summarization, bug detection). Future work should explore additional languages, tasks, and prompt formulations to establish broader applicability.

\section{Conclusion}
This paper examined system prompts as a controllable, first-class factor in instruction-tuned code generation, and evaluated their effects under a unified design that varies prompt specificity, prompting strategy, language, and decoding regime. Concretely, we constructed five progressively more information-dense system prompts and paired them with three prompting strategies (zero-shot, 3-shot fixed, and 3-shot retrieval-based) over Java and Python, reporting functional correctness via Pass@1 and Pass@5 under $T = 0$ and $T = 1$. This yields a comprehensive evaluation matrix of \textbf{360} configurations, enabling controlled comparisons that isolate the interaction between prompt specification and contextual evidence. 

Across models, our results show that prompt specificity does not consistently improve performance; instead, its impact is mediated by both the availability of in-context evidence and the programming language. For the general-purpose model, system prompts with higher specificity act in a strongly interaction-dependent manner: specificity tends to help most when paired with stronger in-context evidence, and retrieval-based prompting provides the most reliable foundation for these gains. Java consistently benefits from increased specificity once relevant examples are available, while Python benefits less consistently and can exhibit counterproductive behavior when added constraints restrict useful exploration under diverse sampling. These observations motivate treating prompt specificity as a targeted design choice rather than a universal improvement, and motivate reporting system prompts as part of the experimental specification rather than treating them as incidental scaffolding. 

For code-specialized instruction-tuned models, we find systematic but non-uniform prompt effects that are structured by capacity and language. A first trend is scale-dependent sensitivity: higher-capacity code models exhibit larger behavioral shifts as prompt specificity increases, indicating that tightly constrained prompts more strongly steer what these models treat as the dominant decision signal during generation.  A second trend is language-conditioned volatility: Java shows condition-dependent prompt effects and clearer strategy-driven regime shifts, whereas Python remains comparatively compressed and stable across many settings. Notably, even when multiple attempts are allowed under stochastic decoding, Pass@5 does not eliminate strategy-driven instability in Java, and prompt preferences remain capacity-conditioned: smaller models repeatedly favor structurally grounded prompts, while larger models benefit from more articulated constraints whose effectiveness depends on whether exemplars are absent or retrieval-aligned. 

Finally, we observe a scale-conditioned inversion for code-specialized models in Java: as model capacity increases, the zero-shot setting more often yields the strongest performance, while adding demonstrations can be counterproductive. In particular, for Qwen-32B and Qwen-7B, introducing 3-shot fixed examples triggers substantial degradation and sharp instability in Java unless the system prompt is sufficiently constraining to re-anchor generation, and even then the recovered performance does not reach the zero-shot peak. This suggests that larger code-specialized models already encode strong Java-specific priors for producing concise, testable methods, and unfiltered examples may introduce distributional drift or competing conventions that interfere with those priors rather than providing useful additional guidance.

\section{Data and Code Availability}
Code, results, scripts, and files used/generated for the investigation have been publicly released in our replication package at: \url{https://github.com/ChrisCheng816/SystemPrompt}.


\bibliographystyle{ACM-Reference-Format}
\bibliography{main}
\end{document}